\documentclass[a4paper,11pt]{article}
\pdfoutput=1
\usepackage{geometry}
\usepackage{jcappub}
\usepackage[T1]{fontenc}
\usepackage{multirow}
\usepackage{array}
\usepackage{gensymb}
\usepackage{xcolor}
\usepackage{amsmath}
\usepackage{xcolor}
\newcommand{\red}{\textcolor{black}}
\newcommand{\blue}{\textcolor{black}}
\newcommand{\purple}{\textcolor{black}}

\title{\boldmath Partial tidal disruption of White Dwarfs in off-equatorial orbits around Kerr black holes}
\author[a]{Aryabrat Mahapatra,}
\author[a]{Adarsh Pandey,}
\author[a]{Debojyoti Garain}
\author[a]{and Tapobrata Sarkar}
\affiliation[a]{Department of Physics, Indian Institute of Technology Kanpur,
	Kanpur 208016, India}
\emailAdd{aryabratm, adarshp, dgarain, tapo@iitk.ac.in}

\abstract
{We present the results of a suite of numerical simulations using smoothed particle hydrodynamics 
to study partial tidal disruption events (TDEs) of white dwarfs (WDs) in off-equatorial orbits in 
intermediate mass spinning (Kerr) black hole backgrounds. We carry out this analysis for both parabolic and eccentric WD orbits and also 
take into account possible initial WD spins. Our objective here is to quantify the differences in variables like the mass of the self-bound core, 
the peak fallback rate of debris and gravitational wave signature in off-equatorial orbits compared to equatorial ones. The analysis is carried out using a hybrid numerical scheme, one which involves integrating the exact Kerr geodesics while adopting a Newtonian formalism for the stellar 
fluid dynamics, justified by our choice of simulation parameters. We find that the physics of TDEs in off-equatorial orbits
present several interesting and novel features due to black hole spin, which in some cases enhances when coupled with the rotation 
of the WD. However, numerical values of observable quantities in TDEs involving off-equatorial orbits cannot 
possibly distinguish between such orbits 
from equatorial ones. We further comment on the genericness of our results and argue that these should extend to a general TDE scenario involving
a spinning BH.}
\begin{document}
\maketitle
\flushbottom

\section{Introduction} \label{sec:intro}
\subsection{Motivation and summary}\label{motivation}
The study of intermediate mass black holes (IMBHs) \citep{Volonteri,Greene} has become an important focus of research
in astrophysics. The main reason for this is that, it is by now believed that IMBHs
maybe the “missing link” between the observationally more well established solar mass
($\sim 3 - 100 M_{\odot}$) BHs and supermassive ($\gtrsim 10^6 M_{\odot}$) black holes (SMBHs). While evidence 
for these two classes of BHs is fairly abundant \citep{Cherepashchuk, KormendyHo}, literature on the detection of IMBHs is relatively rarer, with only
a handful being reported till date, see e.g., \citep{IMBH0, IMBH1, IMBH2, IMBH3, IMBH4, Cao2023}. The study of 
tidal disruption events (TDEs) have a natural role to play here, as TDEs are widely considered 
to be “smoking gun” signatures of BHs. TDEs occur when the self-gravity of a star is overwhelmed by strong non-local gravity of 
a massive object like a  BH, thereby revealing important properties of both the stellar object as well as the BH. The disruption can be
either partial with a surviving self-bound stellar core, or total (full), where the entire stellar material is disrupted. Spinning
(Kerr) BHs assume natural significance in this context, since most confirmed BH sources have an associated 
spin, and thus the importance of understanding TDEs in generic Kerr IMBH backgrounds cannot be over-emphasised. 
Now Kerr black holes being axially symmetric, provide for important 
departures in possible stellar trajectories when compared to the spherically symmetric Schwarzschild BHs. To wit, in the latter case, 
due to this spherical symmetry, any stellar orbit is effectively an equatorial one, a fact that is no longer valid for Kerr BHs. 

The above remarks motivate the need to understand the physics of TDEs in generic off-equatorial inclined stellar orbits in 
Kerr IMBH backgrounds. \blue{Full TDEs from Kerr SMBHs for initially non-rotating polytropic solar mass stars 
in off-equatorial parabolic and eccentric orbits have been studied before, see for example \cite{Haas, Hayasaki2016, LiptaiPrice}.
Here, we study partial TDEs of spinning white dwarfs (WDs) by Kerr IMBHs in such orbits}.
We consider a $0.5M_{\odot}$ WD with a spin period of $5$ minutes in the background of a $10^4M_{\odot}$ Kerr IMBH. The 
pericenter distance (i.e., distance of closest approach) of the center of mass of the WD is fixed at $25$ times the
gravitational radius. As we will explain in details later, these parameters are chosen to validate the applicability 
of our numerical scheme, and serves to illustrate important physics, with sufficient accuracy.  
As is commonly used in similar studies, here we use a numerical procedure based on smoothed particle hydrodynamics 
(SPH) that integrates the Kerr geodesic equations while stellar self gravity is treated in a Newtonian fashion.

Our results in this paper gives a comprehensive analysis of partial TDEs of WDs in the background of Kerr IMBHs. 
We find that the physics of TDEs indicate interesting subtleties in off-equatorial orbits when
compared to equatorial ones, especially when stellar spin is included in the analysis. 
We quantify the impact of the off-equatorial orbits and BH spin parameters on variables 
such as mass disruption from the WD, fallback rates of the debris that is bound to the BH after the disruption, 
and gravitational wave (GW) amplitudes. 
We find however, that using currently established observables, it may not be possible to observationally distinguish between 
partial TDEs corresponding to off-equatorial orbits compared to equatorial ones. As we establish in sequel, 
this is due to the fact that variables in off-equatorial
orbit TDEs always tend to have intermediate values between equatorial orbit ones. The results of off-equatorial TDEs 
might thus be degenerate with those involving equatorial orbits with different BH mass. 
As we will argue in the final section of this paper, this should be a generic 
feature for partial TDEs of WDs for both IMBHs and SMBHs, as borne out by a tidal stress analysis. 

\subsection{Background and review}\label{back}
We now review some of the existing results in the literature that will be relevant to our analysis, based on 
the works of \citep{RR, RR1, Beloborodov, WigginsLie, Ivanov, Ishii, 
Ferrari, Kesden1, SinghKesden, Haas, Hayasaki2016, 2017MNRAS.469.4483T, LiptaiPrice, Banerjee, GaftonRosswog, Jankovic}. 

\blue{In \cite{Kesden1}, the authors studied TDEs by analytic methods, following \cite{Beloborodov}, and showed that the peak accretion 
rate of tidally disrupted debris was achieved when BH spin is anti-aligned with the stellar orbital angular momentum. 
A more recent analysis of TDEs in off-equatorial orbits, via loss cone theory, appears in \cite{SinghKesden}. 
In the context of IMBHs, the work of \cite{RR} performed one of the first numerical analysis of full disruption of polytropic 
solar mass stars in the background of non-rotating IMBHs. Early 
numerical studies involving Kerr IMBHs and WDs appeared in \cite{RR1},\cite{Haas}. 
The work of \cite{RR1} discusses the phenomenon of nuclear burning that can be triggered in WDs by tidal effects of the IMBH in 
full TDEs, while \cite{Haas} also considered similar TDEs leading to full disrpution of solar 
mass WDs in Kerr IMBH backgrounds, and showed that the dynamics of the disruption process depends strongly
on BH spin.}

\blue{Full TDEs involving solar mass polytropic stars and Kerr SMBHs, have been discussed in 
a number of notable works, and we list some of them. The authors of \cite{Hayasaki2016} studied the effects 
of circularisation of debris from stars using post-Newtonian corrected SPH, and 
showed how this process is crucially dependant on the efficiency of radiative cooling, as well as BH spin. Further, \cite{LiptaiPrice} 
used a relativistic version of SPH to illustrate the process of disc formation in such TDEs, and
\cite{GaftonRosswog} details the effects of BH spin and the impact parameter on the outcomes
of these events. }

\blue{Here we consider both parabolic and eccentric orbits of spinning WDs, although we will focus on parabolic
orbits to understand physical observables. It will be prudent to recall here that it is well known that properties of TDEs of stars 
in eccentric orbits might be substantially different from those in parabolic (or nearly parabolic) ones. For example, for full
disruption of stars in parabolic orbits, about half of the stellar debris remain gravitationally bound to the BH, and eventually falls
into it after circularisation. On the other hand, the detailed analysis of \cite{elliptical1} showed that for TDEs in eccentric
stellar orbits, the situation changes drastically, and that the stellar debris might be almost entirely gravitationally bound to the BH,
resulting in important differences in associated observables. 
Further, \cite{elliptical1} also pointed out that for deep encounters of stars in elliptical orbits, relativistic precession may cause
multiple crossing of stellar debris resulting in energy dissipation that leads to faster circularisation. Subsequently, \cite{elliptical2}
further investigated such relativistic effects on debris circularisation, and  emphasised the effects of self intersection of
stellar debris on this process. More recently, \cite{elliptical3} has studied the mass fallback rates of tidally disrupted stars
both in marginally bound and unbound orbits, and showed how these can differ substantially. Note that the works of 
\cite{elliptical1,elliptical2,elliptical3} discussed TDEs of initially non-rotating solar type stars in deep encounters with non-rotating SMBHs.}

\blue{We also recall that the difference between full and partial TDEs have been studied in several important works. Physical features
of TDEs such as the peak fallback rate of stellar debris and its late time slope were characterised in \cite{GRR}, where it was found
that this slope can be close to $-2.2$ for partial disruptions whereas for fully disrupted stars, it is close to $-5/3$, as predicted
from an older Newtonian analysis of \cite{Rees, Phinney}. 
The distinction between full and partial disruptions was further studied by \cite{Mainetti}, who computed
a critical value of the impact parameter below (above) which a polytropic star would be partially (fully) disrupted.
Thereafter, \cite{CN} argued that partial disruptions by SMBHs always produce a mass fallback rate whose late time
slope is $-2.25$, independent of the mass of the remnant self-bound core, and this was numerically verified by \cite{Miles}. 
This result was however contradicted in \cite{2024ApJ...967..167G}, where the authors reported a strong dependence of 
this slope on the mass of the core, for partial TDEs of IMBHs with WDs. 
}

\subsection{Tidal disruption of spinning WDs by Kerr IMBHs}

To further motivate the scenario that we study in this paper, we recall that from the mass-radius relation 
of Carbon-Oxygen WDs \citep{1983bhwd.book}, it can be shown that these can be tidally disrupted either partially or fully only by IMBHs :
SMBHs will, in general, swallow the WD as a whole \citep{Jonker}. Since WDs also have a sufficiently accurate equation
of state, these thus provide an attractive possibility of the study of Kerr IMBHs via TDEs. Here
we study such TDEs by a suite of numerical simulations using SPH. We use a code 
that has been developed by us \citep{2023MNRAS.522.4332B, 2024ApJ...967..167G, 2024arXiv240117031G} and has
been extensively compared and tested with the results available via the publicly available code PHANTOM \citep{2018PASA...35...31P} which
has inspired an initial version of our code. We use a hybrid
formalism for our study, where the stellar particles follow geodesic trajectories in the Kerr IMBH background, while the 
hydrodynamics of the star are treated in a Newtonian manner. The stellar pericenter distance has been suitably chosen so that this 
last assumption for WDs is justified \citep{2024arXiv240117031G}, see subsection \ref{simulation} for further discussion on
this issue. 


To justify our choice of the BH and WD masses that we have indicated in section \ref{motivation}, we first record the formula of the tidal
radius \citep{1975Natur.254..295H} 
\begin{eqnarray}
    r_t \thickapprox R_{\star}\left(\frac{M} {M_{\star}}\right)^{1/3} ~,
\label{equation1}
\end{eqnarray}
where $M$ is the mass of the BH, and $M_{\star}$, $R_{\star}$ being the mass and radius of the stellar object, respectively. A star which approaches
the BH and comes within $r_t$ is tidally disrupted. In practise, this formula is approximate, and there are pre-factors due to stellar fluid dynamics,
stellar spin etc. but this will not affect our qualitative discussion here. 
\begin{figure}[h]
\centering
\includegraphics[width=.6\textwidth]{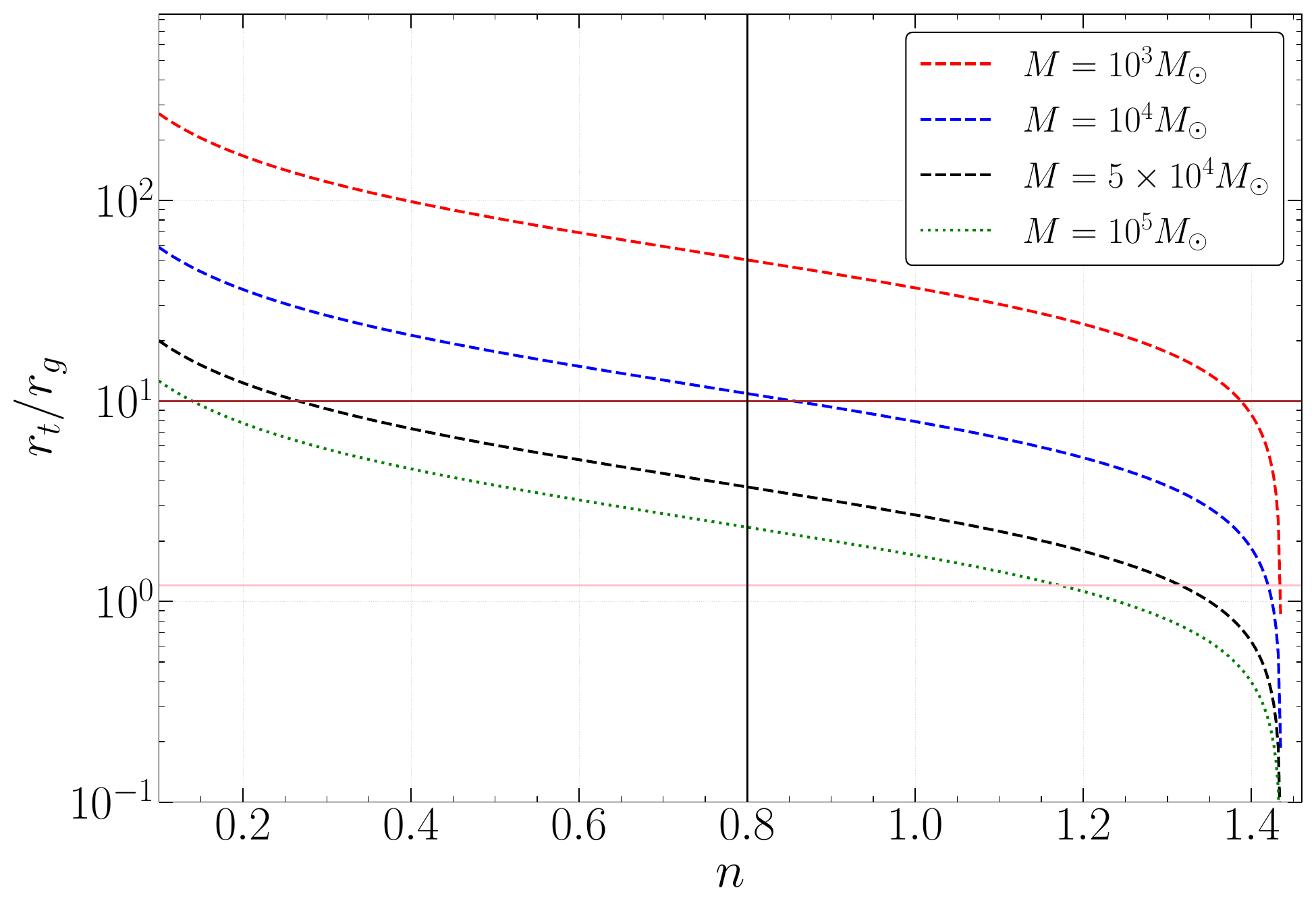}
	\caption{The tidal radius $r_t$ plotted as a function of $n$ where the WD mass $M_{\rm WD}=n\times M_{\odot}$. The curved lines from 
	top to bottom indicate BHs of $10^3,10^4,5\times 10^4$ and $10^5$ solar masses. The brown and pink horizontal lines indicate
	$r_t=10$ and $1.2$ respectively, while the vertical line indicates $n=0.8$.}
    \label{figzero}
\end{figure}

Using the well known mass-radius relation for Carbon-Oxygen WDs, given in e.g. 
Equation 1 on page 39 in the article by Maguire et al. in \cite{Jonker}, we obtain Figure \ref{figzero}, where we plot $r_t/r_g$ (where 
$r_g = GM/c^2$ is the gravitational radius of the BH) as a function of the WD mass $M_{\rm WD}=nM_{\odot}$, with
the Chandrasekhar mass given by $1.435M_{\odot}$. 
In this figure, the pink horizontal line indicates the outer event horizon of a BH with dimensionless spin parameter $a^\star=0.98$, below which
WDs are captured as a whole rather than being tidally disrupted. 
From Figure \ref{figzero}, we see that for $M\gtrsim 5\times10^4M_{\odot}$, most TDEs will be dominated by relativistic effects, assuming that such 
relativistic effects become important for $r_t\leq 10r_g$ \citep{2019GReGr..51...30S}, shown by the brown horizontal line if Figure \ref{figzero}.
On the other hand, for lighter IMBHs with $M\lesssim 10^4M_{\odot}$, using the fact that
most known WDs have masses $\lesssim 0.8M_{\odot}$ \citep{Kepler} indicated by the black vertical line in Figure \ref{figzero}, we see that a majority of 
TDEs of WDs with such IMBHs will be 
away from the relativistic regime. Our choice of parameters in this paper $M_{\rm WD}=0.5M_{\odot}$ corresponding to a radius 
$R_{\rm WD}=0.0141R_{\odot}$ and $M=10^4M_{\odot}$ is indicative of
this regime and we can reasonably expect that our results would be generic for a majority of TDEs involving WDs and lighter 
mass IMBHs. 

This paper is structured as follows: Section \ref{sec:methodology} discusses the methodology to simulate the tidal disruption in off-equatorial places around the Kerr BH using the SPH algorithm and the simulation parameters used in this work. 
Following this, in Section \ref{sec:results}, we discuss the results obtained using our SPH code. 
Finally, Section \ref{sec:conclusion} concludes this paper with discussions and a summary of the results obtained. There are two appendices 
in this paper. Appendix \ref{app:A} contains some verifications of the relaxed WD configurations used here. Appendix \ref{app:B} shows
the convergence of our numerical results. 

\section{Methodology} \label{sec:methodology}
In this section, we discuss the methodology to simulate the tidal disruption of a stellar object in the off-equatorial orbit around a Kerr BH. We employ the SPH algorithm to simulate the encounter between a BH and a stellar object.
SPH is a Lagrangian-based method that conserves energy, linear momentum, and angular momentum. In SPH, the fluid star is discretized into a collection of particles known as SPH particles, each possessing properties such as density, pressure, internal energy, velocity, etc. These 
are smoothed using the $\rm M6$ kernel function. To efficiently compute fluid properties and accelerations, we utilize the binary tree algorithm, as outlined in \cite{2011MNRAS.418..770G}. The tree-accuracy parameter is set to $0.5$, which controls whether a distant node can be treated as a multipole source of gravity or further resolved into its descendants, for a given node. We use the standard SPH artificial viscosity term with the parameters $\alpha^{\mathrm{AV}} = 1.0$ and $\beta^{\mathrm{AV}} = 2.0$, and apply the Balsara switch to reduce viscosity in shear flow, following \cite{1995JCoPh.121..357B}. The SPH fluid equations are integrated using the Leapfrog integrator: a time-reversible and symplectic integration scheme, following \cite{2005MNRAS.364.1105S},\cite{2003AcNum.12.399H}. Global time-stepping is used for the evolution of the system, which ensures the minimization of numerical errors as compared to individual time-stepping. \\

Here, we extend the code used by \cite{2024arXiv240117031G}, where the authors simulate tidal interactions in an equatorial plane to study the effects of BH spin and stellar spin on the observables related to the TDE. 
To do this, we need to fix the initial conditions, i.e., the initial values of the coordinates $r_0, \theta_0, \phi_0$ (we use Boyer-Lindquist
coordinates, see Equation \ref{KerrMetric} below) 
and the initial velocities 
$\dot{r}_0$, $\dot{\theta}_0$, $\dot{\phi}_0$ of the center of mass of the stellar object. For our simulations, we set
$r_0=5 r_t$ from the BH positioned at the origin, $\theta_0=\theta_a$ and $\phi_0=0$, where $r_t$ is the tidal radius, and $\theta_a$ is the initial angle between the inclined orbit and the BH spin direction. Now, to obtain the initial velocities, we need to specify the CM trajectory of the stellar object, which requires the specific energy $\mathcal{E}$, the specific angular momentum $l_z$ (along the BH spin direction), and the Carter's constant $\mathcal{Q}$ \citep{1968PhRv..174.1559C}.
In general, for a given type of orbit, it is also necessary to specify values of the turning points $r_a$, $r_p$, and $\theta_a$, where $r_a$ and $r_p$ are the apocentre and pericentre of the orbit, respectively. 
The pericentre $r_p$ of the orbit is determined using the tidal radius $r_t$ and impact parameter $\beta$ by the relation $r_p = r_t/\beta$, where the tidal radius is given by 
Equation \ref{equation1}. Once we obtain $r_p$, the apocentre $r_a$ is obtained using $r_p$ and eccentricity $e$ of the orbit, via $r_a = r_p(1+e)/(1-e)$. However, specifying $r_a$ and $r_p$ alone is insufficient to determine the conserved quantities in the Kerr geometry for an inclined orbit, as we discuss below.

The Kerr space-time geometry in Boyer-Lindquist (BL) coordinates $(t,r,\theta,\phi)$ is expressed by the following metric:
\begin{eqnarray}
    ds^2 &=& -\left(1-\frac{2GMr}{c^2\Sigma}\right)c^2dt^2 - \frac{4GMra\sin^2{\theta}}{c\Sigma}dtd\phi + \frac{\Sigma}{\Delta}dr^2 
    + \Sigma d\theta^2 \nonumber\\
    &+& \left(r^2+a^2+\frac{2GMra^2\sin^2{\theta}}{c^2\Sigma}\right)\sin^2{\theta}d\phi^2 
\label{KerrMetric}
\end{eqnarray}
where \hspace{0.1cm}$\Sigma=r^2 + a^2\cos^2{\theta}$ \hspace{0.1cm}and\hspace{0.1cm} $\Delta=r^2 - 2GMr/c^2 + a^2$. The Kerr parameter is defined as $a=J/Mc$. Here, $J$ and $M$ denote the angular momentum and mass of the Kerr BH respectively, $c$ is the speed of light and $G$ is Newton's gravitational constant. The dimensionless Kerr parameter is given as $a^{\star}$ = $a/r_g$, where
$r_g = GM/c^2$ is the gravitational radius of the BH. It can vary from $-1$ to $1$, making the rotation of BH from retrograde to prograde, respectively.

We pause to mention here that one could equivalently have used Kerr-Schild coordinates to describe the Kerr BH. For TDEs from
Kerr SMBHs, this was studied in \cite{2017MNRAS.469.4483T}, who found almost identical results for both these coordinates, even 
for highly relativistic encounters (see e.g., Figure 9 of that paper). We will thus focus exclusively on BL coordinates, which is 
standard in the literature on TDEs, since in our case TDEs occur at distances large compared to the 
BH event horizon. 

Our procedure now follows \cite{2017MNRAS.469.4483T}, which we review below. 
To explore the tidal interactions in an off-equatorial orbit along a general trajectory, the turning points of the trajectory is determined from $dr/d\tau=0$. 
This relation is equivalent to setting $\mathcal{R}(r)=0$, where $\mathcal{R}(r)$ is expressed using $r_s$ and $c$ as follows:
\begin{eqnarray}\label{eq2}
    \mathcal{R}(r) = \epsilon r^4 + c^4r_sr^3 - c^2(l_z^2 + \mathcal{Q})r^2 + l^2r_sr-a^2c^2\mathcal{Q}
\end{eqnarray}
where $r_s = 2 r_g$ is the Schwarzschild radius, $l^2$ is the square of total magnitude of specific angular momentum and $\epsilon\equiv\mathcal{E}^2-c^4$. It is evident that the quantity $\mathcal{R}(r)$ is a fourth-order polynomial. Since $dr/d\tau$ vanishes at the turning points, we can rewrite Equation \ref{eq2}, corresponding to its four roots as:
\begin{eqnarray}\label{eq3}
    \mathcal{R}(r) = \epsilon (r-r_a)(r- r_p)(r - r_c )(r - r_d)~,
\end{eqnarray}
where $r_c$ and $r_d$ are two additional roots. From Equations \ref{eq2} and \ref{eq3}, we can write the conserved quantities in terms of turning points as:

\begin{equation}\label{eq4}
\begin{aligned}
    &\epsilon = -\frac{c^4 r_s}{r_a + r_p+ r_c+ _d}~,~~
    \mathcal{Q} = \frac{c^2r_sr_ar_pr_cr_d}{a^2(r_a + r_p+ r_c+ _d)}~,~~
    l^2 = c^4\frac{r_ar_p(r_c+r_d)+r_cr_d(r_a+r_p)}{r_a+r_p+r_c+r_d}~,\\
    &\quad \quad l_z^2 = c^2 r_s\frac{a^2(r_ar_p+r_cr_d+(r_a+r_p)(r_c+r_d)-a^2)-r_ar_pr_cr_d}{a^2(r_a+r_p+r_c+r_d)} \, 
\end{aligned}
\end{equation}

Additionally, for the Kerr metric, the conserved quantities are related as $l^2 = c^2\mathcal{Q}+(l_zc-a\mathcal{E})^2$ and from $d\theta/d\tau = 0$, we have $\mathcal{Q}=l_z^2\cot^2{\theta_a}-\epsilon a^2\cos^2{\theta_a}$. These two equations can be used along with the Equation \ref{eq4}, to obtain the following equations for the unknown turning points $r_c$ and $r_d$:
\begin{eqnarray}\label{eqn-6}
   &~& p - a^2 \cos^2{\theta_a}(r_ar_p + r_cr_d +(r_a+r_p)(r_c+r_d)-a^2 \cos^2{\theta_a})=0 \\
    &~&a^2(q - 2r_s)+  r_cr_p(r_s-r_d) + r_dr_s (r_c + r_p) 
    +r_a(r_c +r_d +r_p)r_s - r_a(r_cr_d+r_p(r_c + r_d))^2\nonumber\\
    &~&- 4r_s(q - r_s)(a^4 + r_ar_pr_cr_d a^2(r_cr_d + r_p(r_c+r_d)+r_a(r_c+r_d+r_p)))=0\label{eqn-7}
\end{eqnarray}
where $p=r_ar_pr_cr_d$ and $q=r_a + r_p +r_c +r_d$.

Since most of our analyses involve parabolic trajectories, the same mechanism can be applied to the parabolic limit $r_a \rightarrow \infty$ and $\mathcal{E} = c^2$, that is, $\epsilon = 0$. In this limit, we arrive at a new set of equations, analogous to those found in Equation \ref{eqn-6} and Equation \ref{eqn-7}, to determine  $r_c$  and  $r_d$  for the parabolic trajectory\footnote{For a parabolic trajectory, we proceed by setting $\mathcal{R}(r) = 0$, which corresponds to a third-order polynomial. This can be rewritten as $\mathcal{R}(r) = c^2 r_s (r - r_p)(r - r_c)(r - r_d)$.}. Once we have $r_p$, $r_a$, $r_c$ and $r_d$ for any general trajectory, we can substitute it back into Equation \ref{eq4} to get $\mathcal{Q}$, $l^2$ and $l_z$. Consequently, we can use these conserved quantities in the first integral of motion to obtain the initial velocities. Once we set the coordinates and velocities, these are then transformed into a Cartesian-like coordinate system and velocities, which are more suitable for our simulation, since spherical symmetry is broken in a TDE. 
Finally, with these initial conditions and the Kerr geodesics accelerations, we obtain the time evolution of the stellar object in the presence of a Kerr BH along an off-equatorial trajectory. \\

After specifying the geodesics, we proceed to configure the details of the stellar object, which is modeled as a WD.
We follow the methodology outlined in \cite{2024arXiv240117031G} to obtain the relaxed WD configuration in SPH. Initially, we derive the theoretical density profile \(\rho(r)\) of WD by solving the hydrostatic equilibrium and mass conservation equations \citep{1935MNRAS..95..207C, 1983bhwd.book}, which are then incorporated into the SPH framework. In our simulations, SPH particles are initially arranged in a closed-packed sphere, which is subsequently stretched using the stretch mapping technique \citep{1994MmSAI..65.1013H} to match the theoretical density profile of the WD. This approach reduces computational costs compared to random particle placements. To minimize fluctuations in fluid properties
introduced during stretch mapping, the WD is evolved in isolation, without external forces. The final relaxed state is achieved when the ratio of total kinetic energy and total gravitational energy falls below a tolerance of \(10^{-6}\) and the density profile converges to the theoretical density profile of WD. 
\blue{Additionally, we should also make sure that the virial equilibrium condition is also satisfied during our numerical procedure. We recall that
this is the statement \cite{1983bhwd.book} that in hydrostatic equilibrium, the gravitational potential energy of the star $W$ satisfies
\begin{equation}
W + 3\Pi = 0~,~\Pi = 4\pi\int_0^{R_{\rm WD}}Pr^2 dr~,
\end{equation}
where $P$ is the electron degeneracy pressure inside the WD of radius $R_{\rm WD}$. Following \cite{1983bhwd.book}, the degeneracy
pressure is 
\begin{equation}\label{pwd}
P = \frac{m_e^4 c^5}{\hbar^3}\phi(x)~,
\end{equation}
where, $\phi(x) = \frac{1}{8\pi^2}\left(x(1+x^2)^{1/2}(\frac{2x^2}{3}-1) + \mathrm{ln}[x^2 + (1+x^2)^{1/2}]\right)$.
Here, the Fermi momentum is denoted by $x=p_f/{m_e c}$, where $m_e$ is the mass of the electron and 
$\hbar$ is the reduced Planck's constant. Further, from the formula for number density of electrons at zero temperature denoted 
as $n_e$, the mass density is obtained as $\rho = n_e \mu_e$ where $\mu_e$ (in atomic mass units) is the mean 
molecular weight per electron, one obtains \cite{1983bhwd.book}
\begin{equation}
\rho = \mu_e\frac{m_e^3c^3}{3\pi^2\hbar^3}x^3~.
\end{equation}
Now using the hydrostatic equilibrium condition $\nabla P = -\rho\nabla\Phi$ with $\Phi = Gm(r)/r^2$ with $m(r)$ denoting
the mass of stellar matter up to radius $r$, one can obtain numerical solutions of $x(r)$ and $m(r)$, which can then be integrated
to check the condition for virial equilibrium. In our case, we find that during the numerical process of stellar relaxation,
the deviation from viriality is less than $1\%$, thus confirming that virial equilibrium is reached when the star is
in its final relaxed state.
}

\purple{In Appendix \ref{app:A}, we present the final relaxed state of our SPH WD model, including the mass–radius relation calculated from the final SPH measurements. We also provide the internal profiles of the relaxed configurations, which are compared with the corresponding theoretical results, and we show the relative differences between them. These analyses are done for pure Carbon-Oxygen (C-O) WDs. Moreover, these WD configurations still remain unchange for other compositions, such as pure He and O-Ne-Mg WDs, since, $\mu_e=2$ in all cases. Hence, the solutions corresponding to these WD compositions are identical, leading to the same equilibrium WD configuration.}

Now to construct a uniformly rotating WD, following \cite{2020A&A...637A..61G}, we impose rigid rotation on the previously relaxed non-rotating WD configuration.  To reduce noise in velocity, and achieve equilibrium quickly, velocities are periodically reset to zero in the comoving reference frame, initially at shorter intervals and later at longer intervals. Once the angular velocity settles within $0.1\%$ of the desired angular velocity, the system is allowed to evolves freely without resetting the velocity,
and we check whether the central density, the equatorial and polar radii, and the energies oscillate within $2.5\%$ of their average values. Finally, the 
relaxed rotating WD configurations are considered to form when all these properties remain within this specified tolerance for a sufficiently long duration.
The maximum angular velocity of a star that can be sustained without breaking apart is determined by equating the self-gravitational force with the centrifugal force given by \(\Omega_{\text{br}} = \sqrt{GM_{\star}/R^3_{\star}}\). Here, we will use the dimensionless quantity known as the breakup fraction, defined as $\lambda = \Omega_{\rm WD}/\Omega_{\text{br}}$ where $\Omega_{\rm WD}$ is the angular velocity of the spinning WD.

\subsection{Simulation Parameters}\label{simulation}
In this work, we adopt a wide range of combinations of all parameters required to explore the tidal interaction between a 
rapidly spinning Kerr BH and a spinning Carbon-Oxygen WD. 
We examine a WD orbiting a BH in two configurations: a parabolic orbit with eccentricity $e = 1$ and an elliptical orbit with $e = 0.9$, both with a pericenter distance $r_p \simeq 25 r_g$. In this paper, we consider the impact parameter $\beta = r_t/r_p= 0.7$ that corresponds
to partial disruption scenarios of a WD due to an IMBH. \blue{Note that in our case, full tidal disruption occurs at
$\beta \sim 0.85$, while for $\beta \lesssim 0.6$, the disruption becomes feeble. In order to capture 
the essential physics while safely being in a regime where the hydrodynamics can be considered Newtonian, 
we have chosen $\beta = 0.7$. 
Note that typically in the literature, TDEs with values of tidal radius $r_t \lesssim 10r_g$ are considered
to be considerably influenced by general relativistic effects \cite{2019GReGr..51...30S},\cite{CN1}. Further, \cite{CN1} predicts
that such relativistic effects will be considerable for $r_p \leq 10r_g$. Here, 
we have $r_t \simeq 18r_g$, and $r_p \simeq 25 r_g$, which are in the range in which
it is reasonable to assume that Newtonian stellar hydrodynamics is a sufficiently robust approximation for our purpose. }

In our case the spin of the WD corresponds to a break-up fraction $\lambda=\pm 0.08$ where the plus (minus) signs denote prograde (retrograde)
motions that are aligned parallel (opposite) to the direction of orbital rotation. This will significantly preserve the spherical shape of the WD before tidal interaction. 
We have verified that the aspect ratio, defined as the ratio of the equatorial to the polar radius of the undisrupted WD
is maintained around $\approx 1.001$. The relative difference in the tidal radius of spinning WDs from its non-spinning counterpart is less than $1\%$ according to the effective tidal radius formula of \cite{2019ApJ...872..163G}.

In our simulation, we always fix the Kerr IMBH spin along the positive or negative $z$ direction. Now, we have a plethora of options for configuring our simulations to take place on 
any (initially) inclined orbit around the Kerr IMBH, with an inclination angle ranging from $\theta_a=0^\circ$ to $\theta_a=180^\circ$, which fixes the 
initial orbital angular momentum (perpendicular to that orbital plane). Once we fix our orbit, it is necessary to set the spin angular momentum of the spinning WD. To draw significant observational consequences, we will only focus on the cases that result in maximum (or minimum) effects on physical variables. 
For this, the initial spin angular momentum of spinning WD is set to be either parallel or antiparallel with its initial orbital angular momentum.

We take the Kerr spin parameter to be $a^{\star}=-0.98$ and $a^{\star}=+0.98$ and run a suite of simulations by varying 
the angle of inclination between  $\theta_a=1^{\circ}$ and at $\theta_a=90^{\circ}$. The angle of inclination 
$\theta_a=0^\circ$ leads to a divergence in the numerical formalism as the determinant of the metric of 
Equation \ref{KerrMetric} diverges there, while $\theta_a>90^{\circ}$ 
is redundant by virtue of axial symmetry. The angles between $\theta_a=1^{\circ}$ 
and $\theta_a=90^{\circ}$ give intermediate results of the effects of tidal disruption between these two cases, and here
we show the simulations for only these two values of $\theta_a$ to contrast the differences in off-equatorial simulations and
equatorial ones.


For our partial disruption scenarios, we identify the self-bound core particles formed after the disruption using an energy-based iterative method following \cite{GRR}. 
The core formed after the partial disruption has high-density, which in SPH results in extremely small time steps, 
that slows down the evolution of the system. 
To efficiently evolve the system, once the core moves away from the BH and its properties are saturated, the core is replaced by a sink particle to 
speed up the simulations, following \cite{2023JCAP...11..062G, 2018PASA...35...31P}. We have numerically verified that the sink placement time does not
affect physical variables like mass fallback rates. In our simulations, we replaced the core with a sink particle at approximately $0.24$ hours.

We modeled our relaxed WD using $5\times10^5$ SPH particles. Furthermore, we have checked the convergence of our results 
with $10^6$ particles. We have shown some key results in Appendix \ref{app:B} that validate this resolution check.

\section{Results} \label{sec:results}

Now we explore the outcomes of the TDE of a WD following an off-equatorial trajectory around a Kerr BH. First, we focus on the mass loss from the WD due to partial disruption, and the distribution of debris in terms of specific energy. Following that, we discuss the impact of tidal interactions on key observables, including fallback rates, kick velocities, and GW emission during pericenter passage.

\subsection{Mass Disruption}
To begin with, we study the effects of orbital inclination on the resulting self-gravitating core and debris mass distribution in the absence of WD spin for a parabolic orbit. This is then compared to a bound interaction, where the WD follows an eccentric orbit $e = 0.9$. Subsequently, we investigate how introducing an initial WD spin influences the core mass fraction (defined as the ratio of the core mass to the initial mass of the WD) due to the coupling between the WD spin, its orbital angular momentum and BH spin.

\begin{table}[htbp]
    \centering
    \begin{tabular}{|c|c|c|c|c|}
        \hline
        \multirow{2}{*}{$e$} & \multicolumn{2}{c|}{Relative difference in $M_{\text{core}}/M_{\text{wd}}$} & \multicolumn{2}{c|}{Relative difference in $M_{\text{core}}/M_{\text{wd}}$} \\ 
        & \multicolumn{2}{c|}{between $\theta_a=1^\circ$ \& $\theta_a=90^\circ$} & \multicolumn{2}{c|}{between BH spins $a^\star=+0.98$ \& $a^\star=-0.98$} \\ \cline{2-5}
        & $a^{\star} = +0.98$ & $a^{\star} = -0.98$ & $\theta_a = 1^\circ$ & $\theta_a = 90^\circ$ \\ \hline
        $1.0$ & 6.42 \% & 11.29 \% & 0.31 \% & 16.16 \% \\ \hline
        $0.9$ & 8.51 \% & 24.04 \%& 0.57 \% & 26.70 \% \\ \hline
    \end{tabular}
    \caption{Comparison of the relative difference in $M_{\rm core}/M_{\rm wd}$ between two inclined orbits for the specified values of $a^{\star}$ and $\theta_a$, in parabolic and eccentric orbits, in the absence of WD spin.}
    \label{table-1}
\end{table}

\begin{figure}[h]
\centering
\includegraphics[width=.45\textwidth]{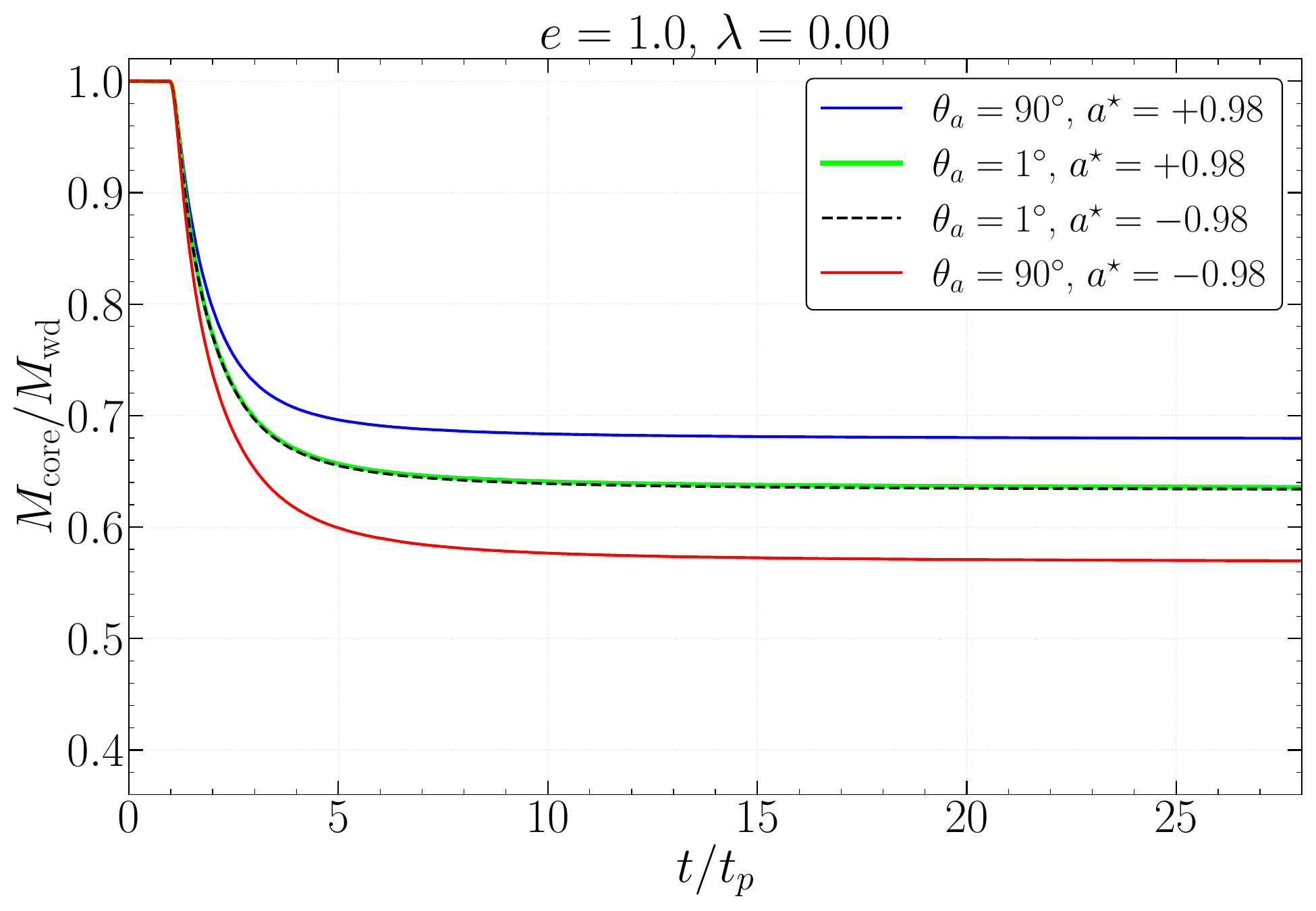}
\hfill
\includegraphics[width=.45\textwidth]{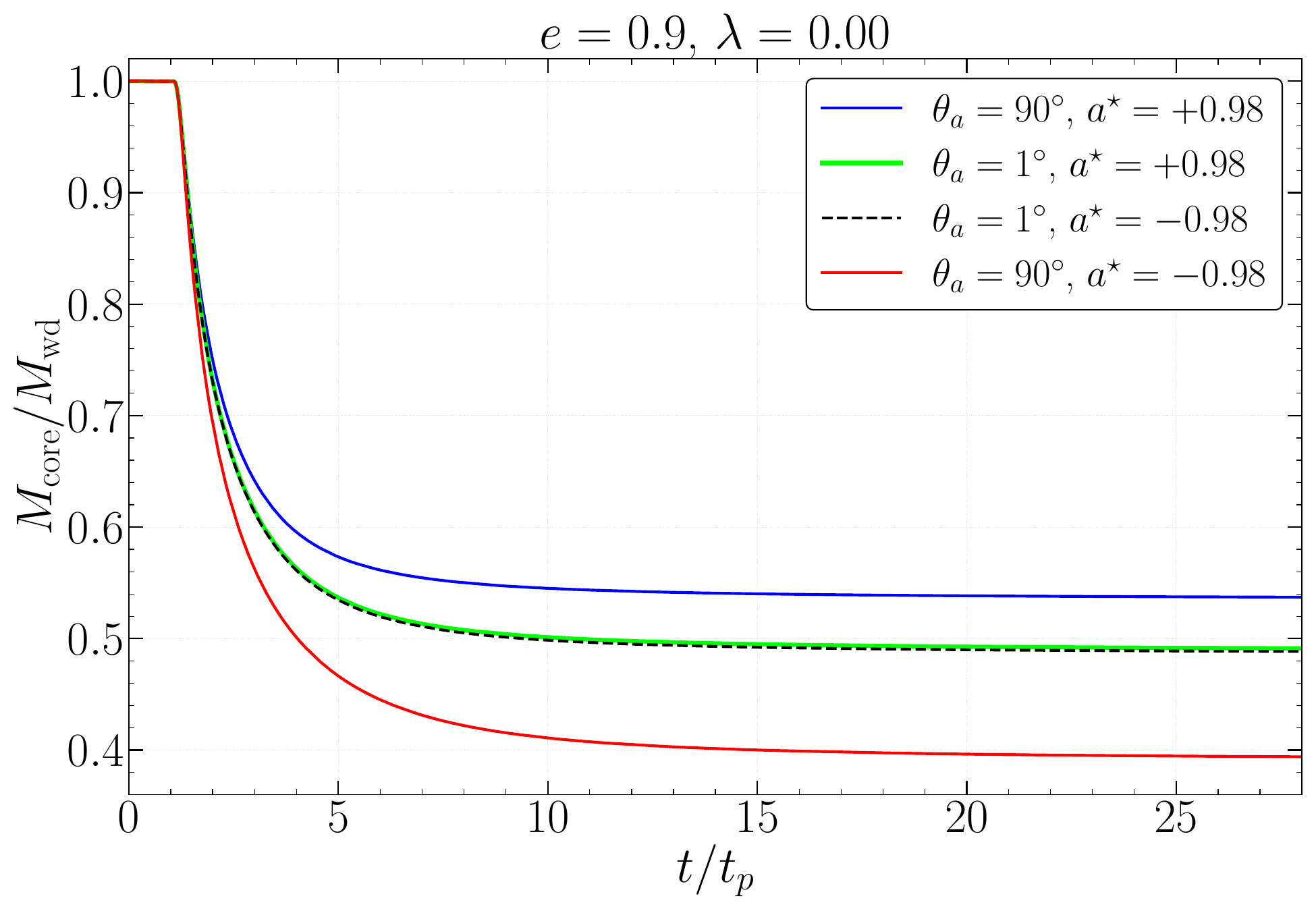}
\caption{The core mass fraction,  $M_{\text{core}} / M_{\text{wd}}$ , is plotted against the normalized time  $t/t_p$  where $t_p\approx28$ sec for the specified values of inclination angle, 
$\theta_a$ and BH spin, $a^{\star}$. \textbf{Left Panel} shows the variation for a parabolic orbit, whereas \textbf{Right Panel} shows the variation 
for an eccentric orbit with  $e = 0.9$.}
 \label{fig-1}
\end{figure}

In the absence of spin of the WD, we illustrate the resulting core mass fraction in Figure \ref{fig-1}. Fixing the BH spin (at $a^{\star}=\pm 0.98$) and considering
inclined orbits at $\theta_a=1^{\circ}$ and $\theta_a=90^{\circ}$, we observe two distinct results in terms of the total mass disrupted from the initial WD. The disruption 
is found to be maximal in the $\theta_a=1^{\circ}$ and minimal in the $\theta_a=90^{\circ}$ for $a^{\star}=0.98$. We then repeat this analysis for a BH spin 
of $a^{\star}=-0.98$. This time, the results are reversed: the disruption is minimal in the $\theta_a=1^{\circ}$ and maximal in the $\theta_a=90^{\circ}$. The relative difference\footnote{The relative difference between equatorial and off-equatorial orbits is computed by taking the equatorial plane,  as the reference. Likewise, the relative difference between BH spins is computed by taking the $a^\star=+0.98$ as the reference.}  of bound core mass fraction between the two inclined orbits are given in Table \ref{table-1}. These results are collectively presented on the left panel of Figure \ref{fig-1}, where the four mass disruption curves are plotted on a normalized mass and time scale. Examining the first row and comparing the first and second columns of Table \ref{table-1} 
shows that a negative BH spin has the most significant effect on the tidal disruption of the WD. 
Additionally, we find that for the $\theta_a = 90^{\circ}$, the relative change in the core mass fraction by changing the BH spin parameter from $a^{\star} = -0.98$ to $a^{\star} = 0.98$ results in significant deviations. 
On the other hand, in the $\theta_a=1^{\circ}$, the curves for both positive and negative BH spins almost coincide, showing little to no distinction 
between the two spins of the BH. 
This distinctive feature of the inclination effect\footnote{We attribute this inclination effect to the fact that the outcome in the $\theta_a = 1^\circ$ is the same for both BH spins, $a^\star = \pm 0.98$. Furthermore, this outcome lies between the distinct outcomes observed for $\theta_a = 90^\circ$, $a^\star = +0.98$ and $\theta_a = 90^\circ$, $a^\star = -0.98$.} is preserved even when the WD follows an elliptical orbit, as shown in the right panel of Figure \ref{fig-1}. However, we observe that the core mass fraction for the elliptical orbit is lower than that of the parabolic orbit when the BH spin parameter and initial inclination angle are the same.
This is likely to happen due to the fact that a WD passing in a bound orbit, will spend a longer duration in the region of strong tidal interaction. 

\begin{figure}[h]
\centering
\includegraphics[width=.45\textwidth]{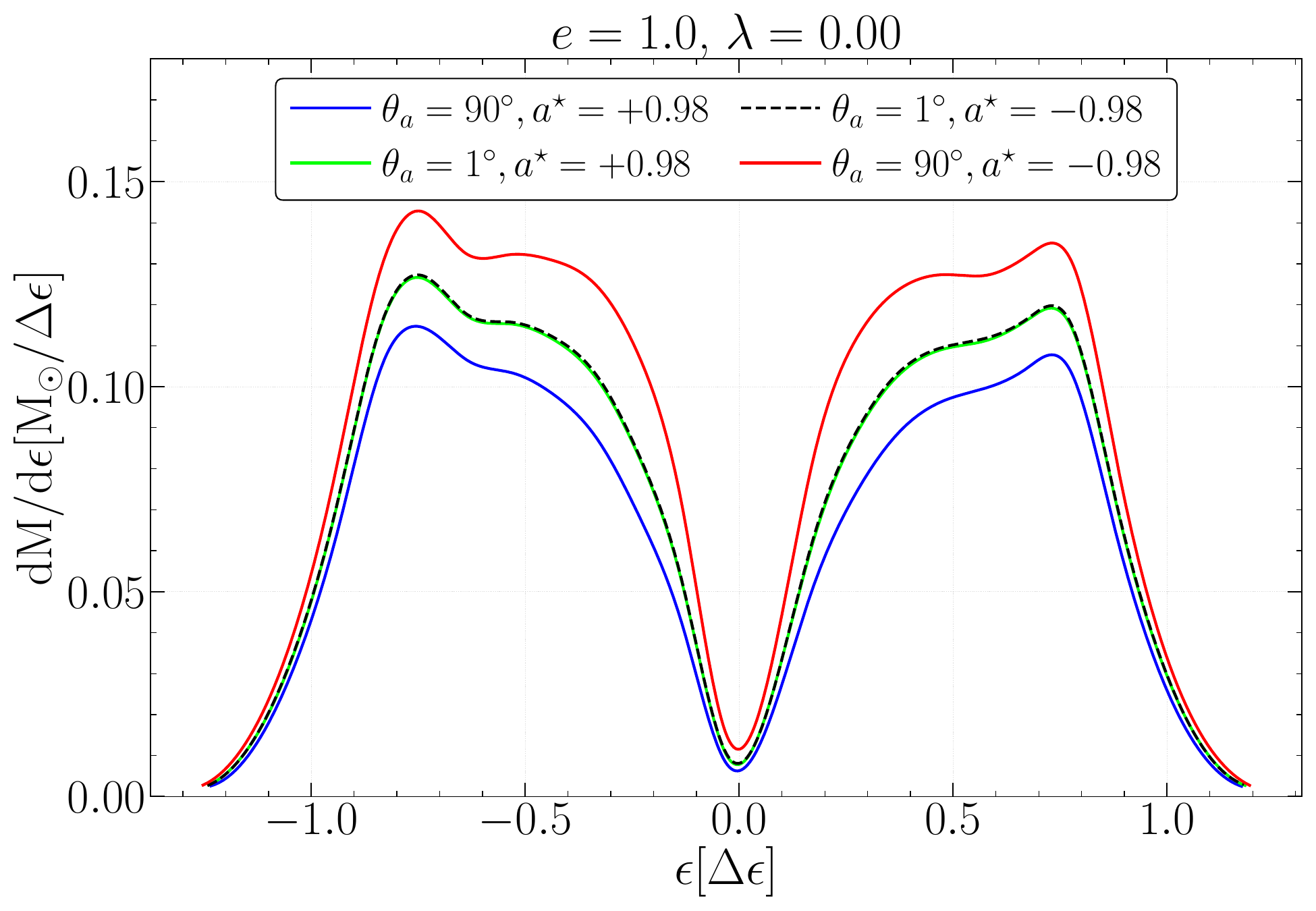}
\hfill
\includegraphics[width=.45\textwidth]{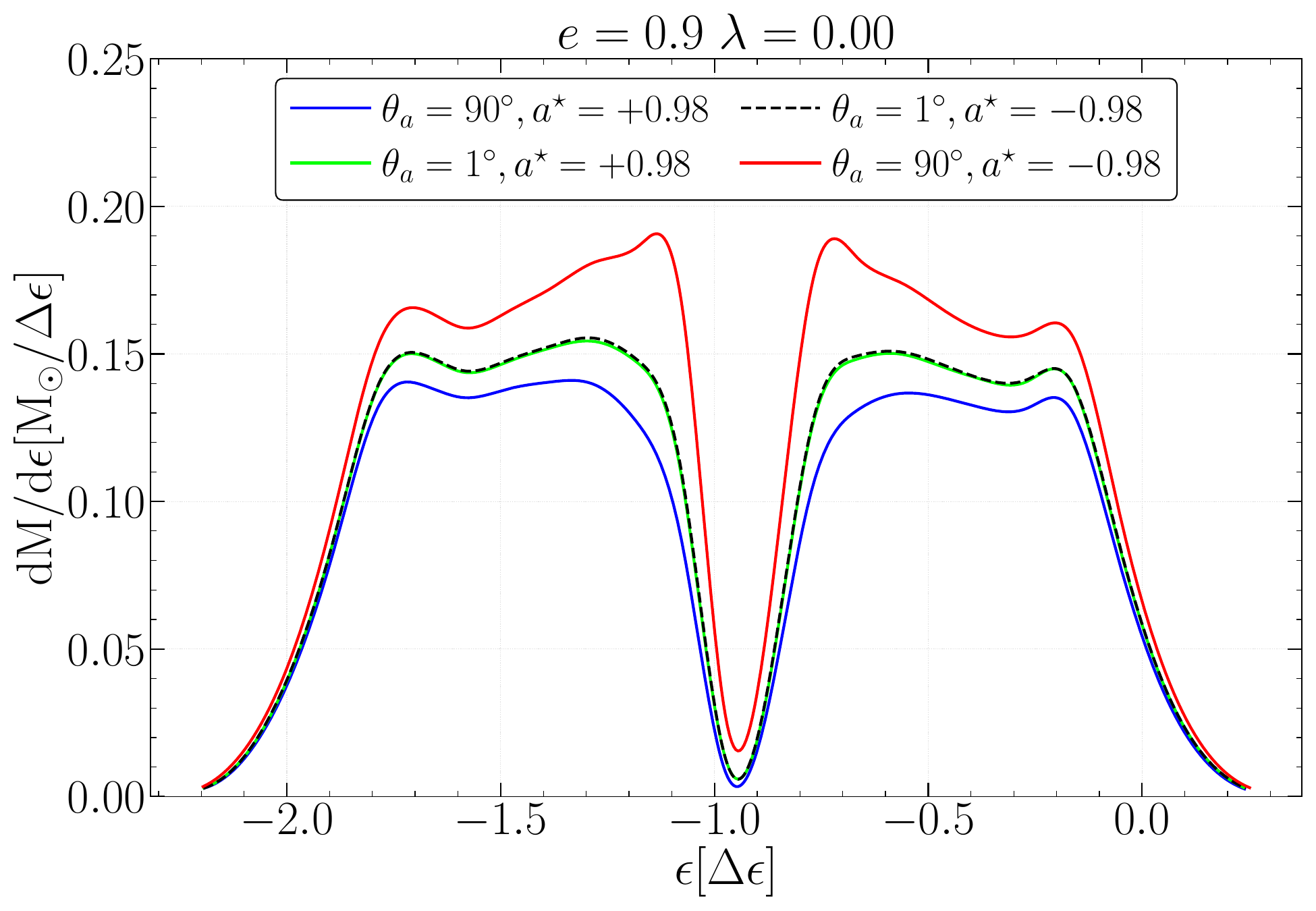}
\caption{ The differential mass distribution, $\text{dM}/\text{d}\epsilon$, is plotted against the normalized specific energy $\epsilon$ for the specified values of inclination angle, $\theta_a$ and BH spin, $a^{\star}$. \textbf{Left Panel} shows the variation for a parabolic orbit, whereas \textbf{Right Panel} shows the variation for an eccentric orbit with $e = 0.9$}
    \label{fig-2}
\end{figure}

Next, we analyze the differential mass distribution of the debris with specific energy for both the parabolic and elliptical orbits. Parabolic orbits lead to the formation of two distinct tidal tails along with the self-gravitating core \citep{2015MNRAS.449..771G, 2014ApJ...782L..13M}, with one tail consisting of part of the debris which is bound to the BH with 
negative specific energy and the other consists of the part which is unbound from the BH with positive specific energy. This is also evident in our calculation of the differential mass distribution of the debris, which is obtained once the core mass saturates to form a self-bound object. In the left panel of Figure \ref{fig-2}, we have computed the differential mass distribution,  $\text{dM/d}\epsilon$, for the debris of both tidal tails of the WD in a parabolic orbit, using the post-disruption snapshot taken at  $t \approx 0.22$ hr (at this instant, no mass has
crossed into the accretion radius of $3r_t$, see discussion in section \ref{observables}). 

The debris differential mass distribution is plotted against the specific energies, measured in units of the spread in specific energy of the debris,  $\Delta \epsilon = G M R_{\text{wd}}/r_t^2$, for four different configurations of BH spins and inclined orbits. The effect of inclination in all these configurations follows a similar outcome, where the curves corresponding to the $\theta_a = 1^\circ$  orbits overlap for both $a^\star = \pm 0.98$. In contrast, the equatorial plane shows different results for $a^\star = \pm 0.98$. In the elliptical orbit case, the differential mass distribution profiles are enhanced and feature higher peaks compared to the parabolic case, reflecting a greater extent of mass disruption, see the right panel of Figure \ref{fig-2}. In this case, we have used the post-disruption snapshot taken at $t\approx0.11$hr, after the saturation of bound core mass and before the accretion of debris. However, we observe that this distribution spread across primarily over negative values of specific energy, since in our case, the tidal tails in elliptical orbits correspond mostly to bound debris. 
It can also be verified from \cite{2018ApJ...855..129H}, that the simulation parameters considered here in our work satisfy the condition  $e^-_{\text{crit}} < e < 1$, where  $e^-_{\text{crit}} = 1 - (2/\beta )q^{-1/3}$  with the mass ratio  $q = M/M_\star$. In our case, with $e^-_{\text{crit}} = 0.89$, the elliptical orbit ($e = 0.9$) is marginally eccentric. As a result, most of the debris remains bound to the BH, with only a small fraction becoming unbound.
In general, tidal disruption phenomena occurring in off-equatorial trajectories are often understood by examining the relative tidal 
strength in various off-equatorial orbits \cite{2024MNRAS.527.6233M, 2012PhRvD..85b4037K, SinghKesden}. Here, we will 
examine test particle trajectories and analyze how their accelerations vary with radial distance, offering a simple and 
intuitive explanation for the inclination-dependent effects.

\begin{figure}[h]
    \centering
    \includegraphics[width=1.0\linewidth]{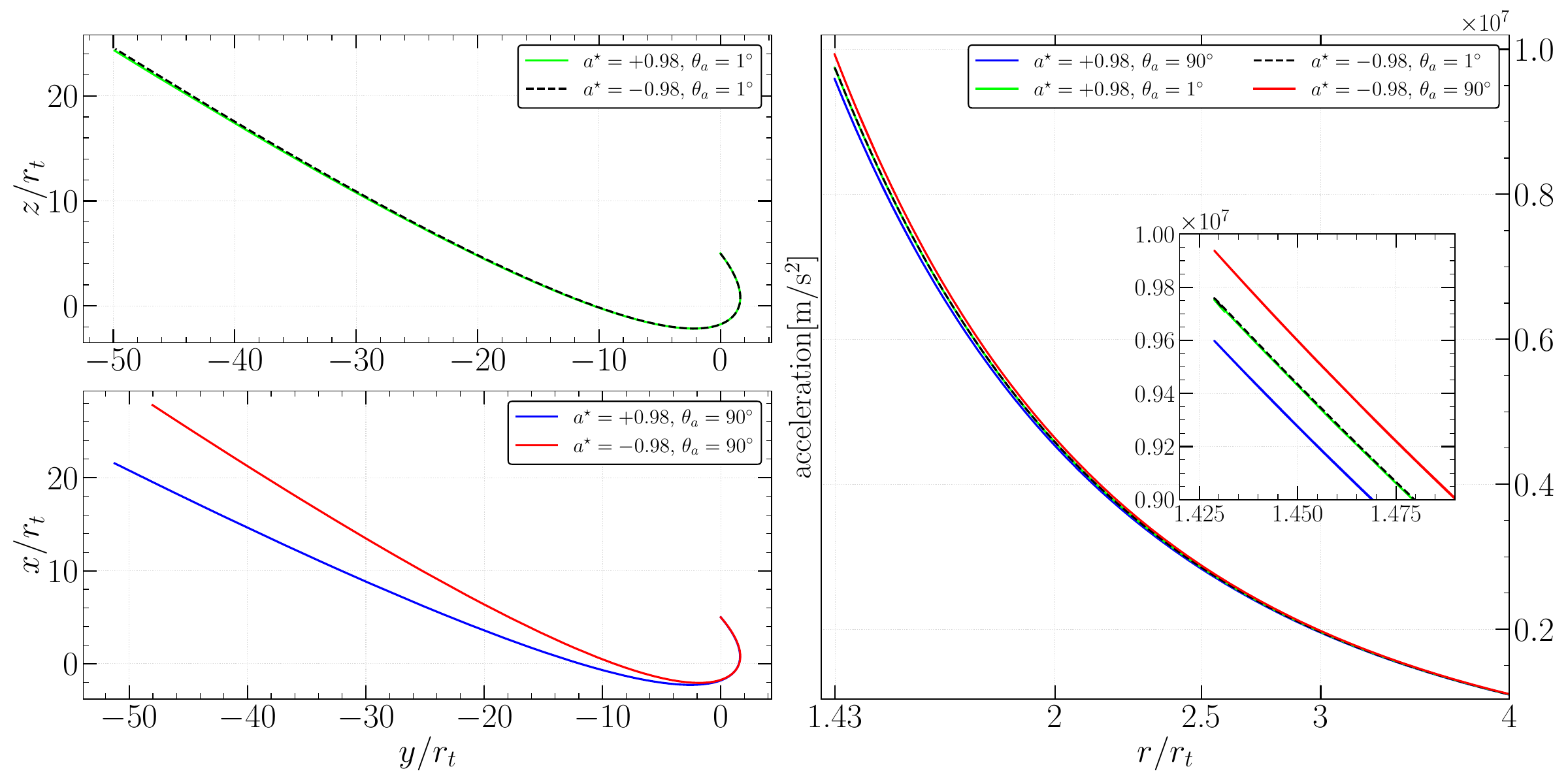}
    \caption{Geodesics and accelerations for a test particle orbiting in an off-equatorial parabolic orbit around a Kerr BH at the origin with spin  $a^{\star}=\pm0.98$. The coordinates and distances are normalized with tidal radius, $r_t$ of the BH. \textbf{Top Left Panel:} The $y$--$z$ projection of the test particle’s trajectory in the inclined orbit $\theta_a = 1^{\circ}$. \textbf{Bottom Left Panel:} The trajectory of the test particle for  $\theta_a = 90^{\circ}$, i.e., on the $x$--$y$ plane (equatorial orbit). \textbf{Right Panel:} The variation of the magnitude of the acceleration with radial distance for all the specified cases.}
    \label{fig-3}
\end{figure}

We consider the same initial conditions as chosen for our simulations for the test particle. In the top left panel of Figure \ref{fig-3}, we show the projection of test particle geodesics on the $y$–$z$ plane, corresponding to motion in the $\theta_a=1^{\circ}$ off-equatorial orbit around a Kerr BH with both spins. Regardless of the BH spin, the trajectories overlap and show little discernible difference due to the spin. The bottom left panel of Figure \ref{fig-3} illustrates test particle geodesics in the $\theta_a=90^{\circ}$ plane, which is confined to the $x$–$y$ plane. In this case, the trajectory for $a^{\star} =-0.98$ brings the test particle into closer proximity after pericenter passage as compared to the $a^{\star}=0.98$ case, due to the larger apsidal precession. Finally, in the right panel of Figure \ref{fig-1}, we present the variation of test particle acceleration as a function of radial distance. It is evident that in the $\theta_a=1^{\circ}$ orbit, the test particle experiences the same acceleration for both BH spins. Additionally, focusing on the region near the pericenter (around $r_p\approx1.43 r_t$), we observe that the test particle in the $\theta_a=90^{\circ}$ orbit, with $a^{\star}=-0.98$, experiences the highest acceleration, while the particle with $a^{\star}=+0.98$ experiences the least. The test particle in the $\theta_a=1^{\circ}$ orbit has an intermediate influence for both BH spins. Using this notion, at the pericenter, we can interpret that WD experiences similar tidal effects in the $\theta_a = 1^\circ$ orbit for both BH spins. However, in the equatorial plane, it experiences the highest and lowest tidal effects for BH spins $a^\star = -0.98$ and $a^\star = +0.98$, respectively. Thus, in the $\theta_a = 1^\circ$ orbit, we observe the same bound core mass fraction for both BH spin values, while in the equatorial plane, the bound core mass fraction varies depending on the BH spin. As a result, the $\theta_a = 1^\circ$ orbit gives rise to the same amount of disrupted mass, consisting of bound and unbound debris, leading to identical debris differential mass distribution profiles for both BH spins. However, in the equatorial orbit, the debris differential mass distribution profile is most enhanced for the negative BH spin due to the higher amount of debris, while the opposite occurs for positive BH spin. 

\begin{figure}[h]
	\centering
	\includegraphics[width=.47\textwidth]{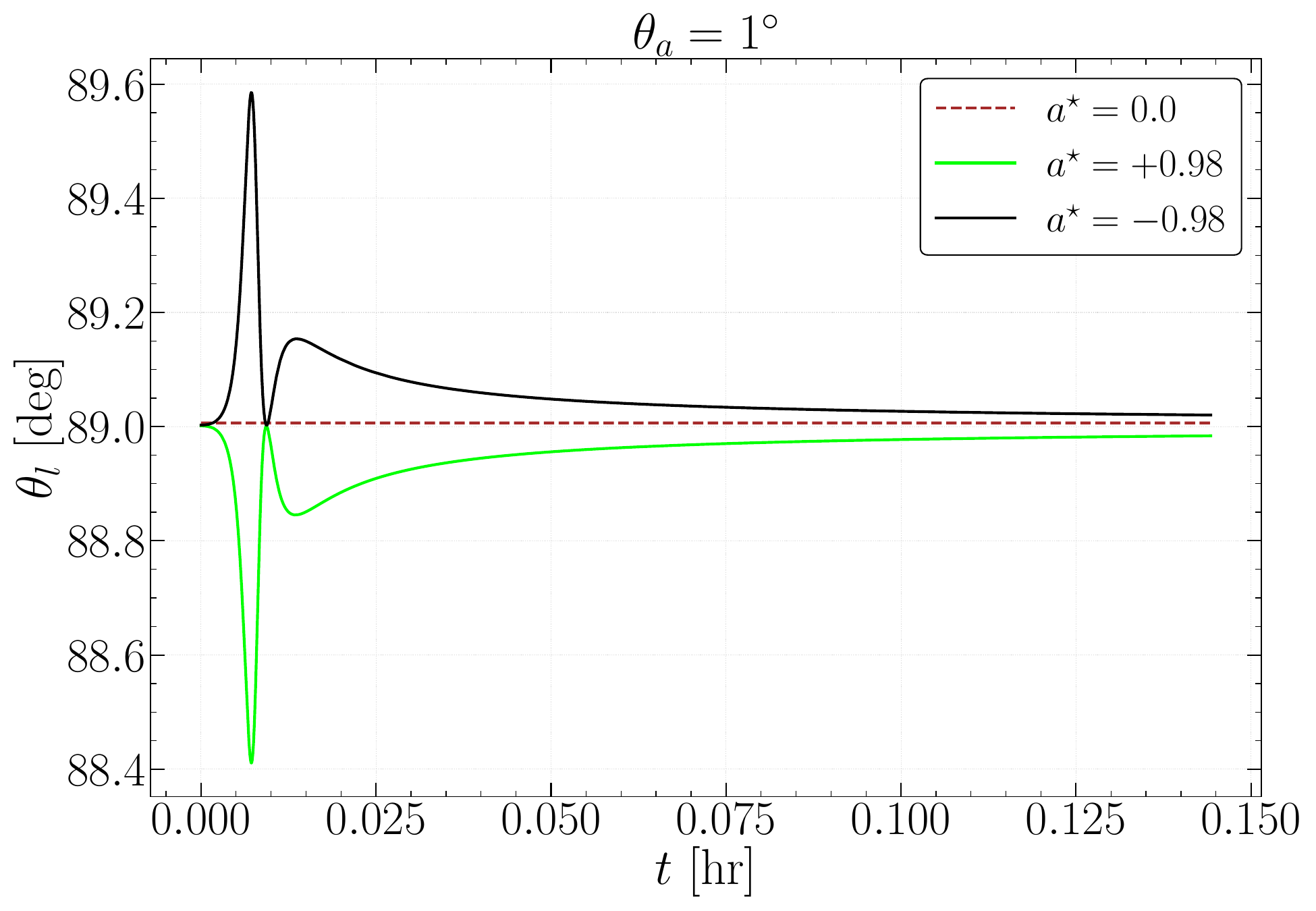}
	\hfill
	\includegraphics[width=.47\textwidth]{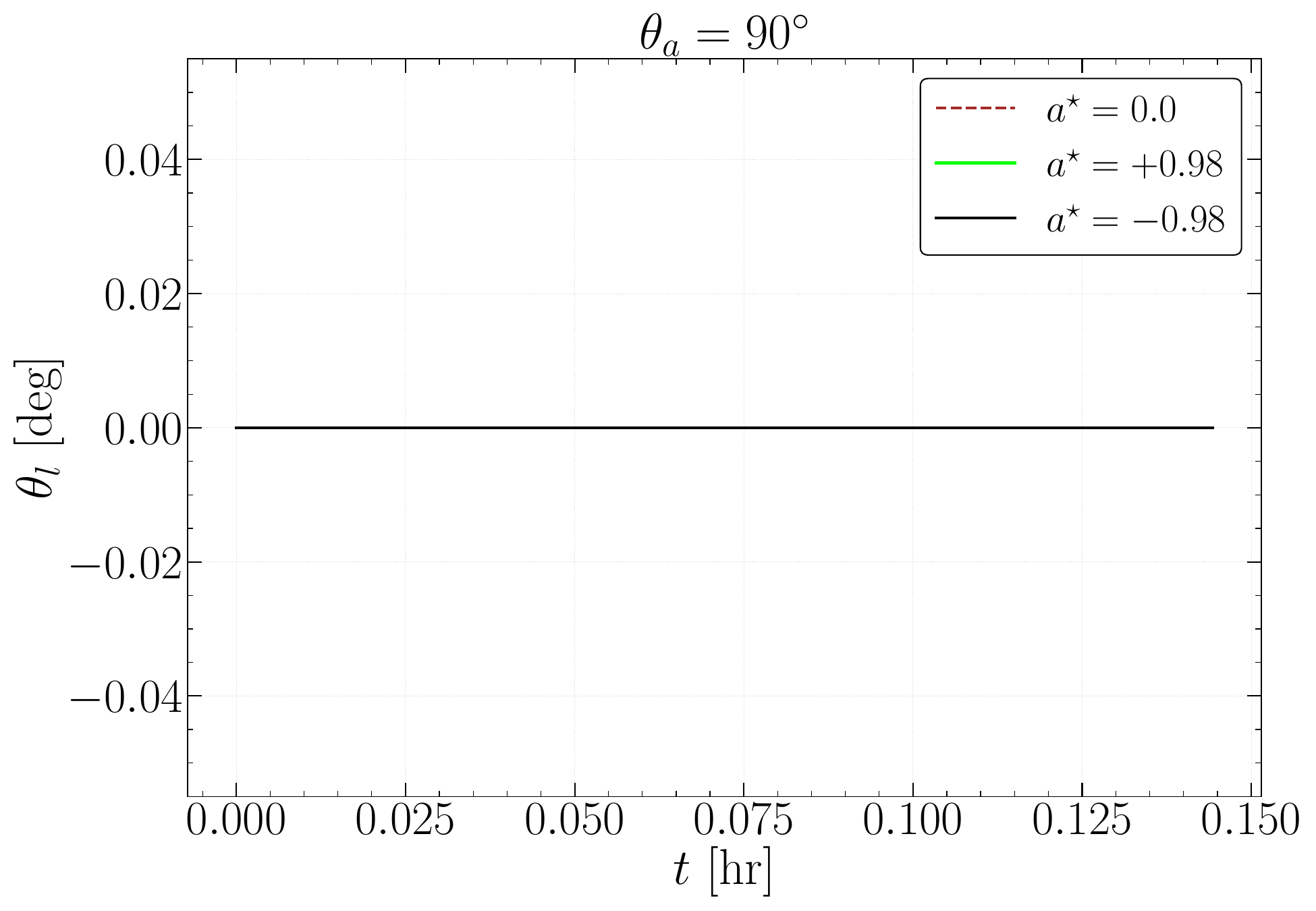}
	\caption{Angle subtended by the orbital angular momentum of a test particle with the $z$-axis. \textbf{Left Panel: }For the $\theta_a = 1^\circ$ case, the orbital motion of the test particle is not strictly confined to a plane around the spinning BH; however, it remains well restricted to $\theta_a=1^\circ$ off-equatorial plane around non-spinning BH. \textbf{Right Panel: }For the equatorial case ($\theta_a = 90^\circ$), the orbital motion lies entirely within the plane for both spinning and non-spinning BH.}
	\label{fig-3.1}
\end{figure}

\red{Using this test-particle dynamics, we also provide a brief demonstration of the Lense Thirring effect in Figure \ref{fig-3.1}. We plot the evolution of the angle $\theta_l$ subtended by the orbital angular momentum of the test particle with respect to the $z$-axis around both spinning and non-spinning BHs. In the left panel of Figure \ref{fig-3.1}, for off-equatorial orbits with $\theta_a = 1^\circ$, $\theta_l$ exhibits a small deviation, with a maximum relative difference of $\approx 0.67\%$ for both $a^{\star} = \pm 0.98$. This relative difference is measured with respect to the $\theta_l$ value for the non-spinning BH, which remains constant at all times. The small deviation observed in the spinning BH cases can be attributed to the effective Lense–Thirring effect in our off-equatorial setup. While, for $\theta_a = 90^\circ$, the dynamics remains confined to the equatorial plane, as shown in the right panel of Figure \ref{fig-3.1}, where $\theta_l$ always stays zero, as expected.}

\begin{figure}[h]
\centering
\includegraphics[width=.45\textwidth]{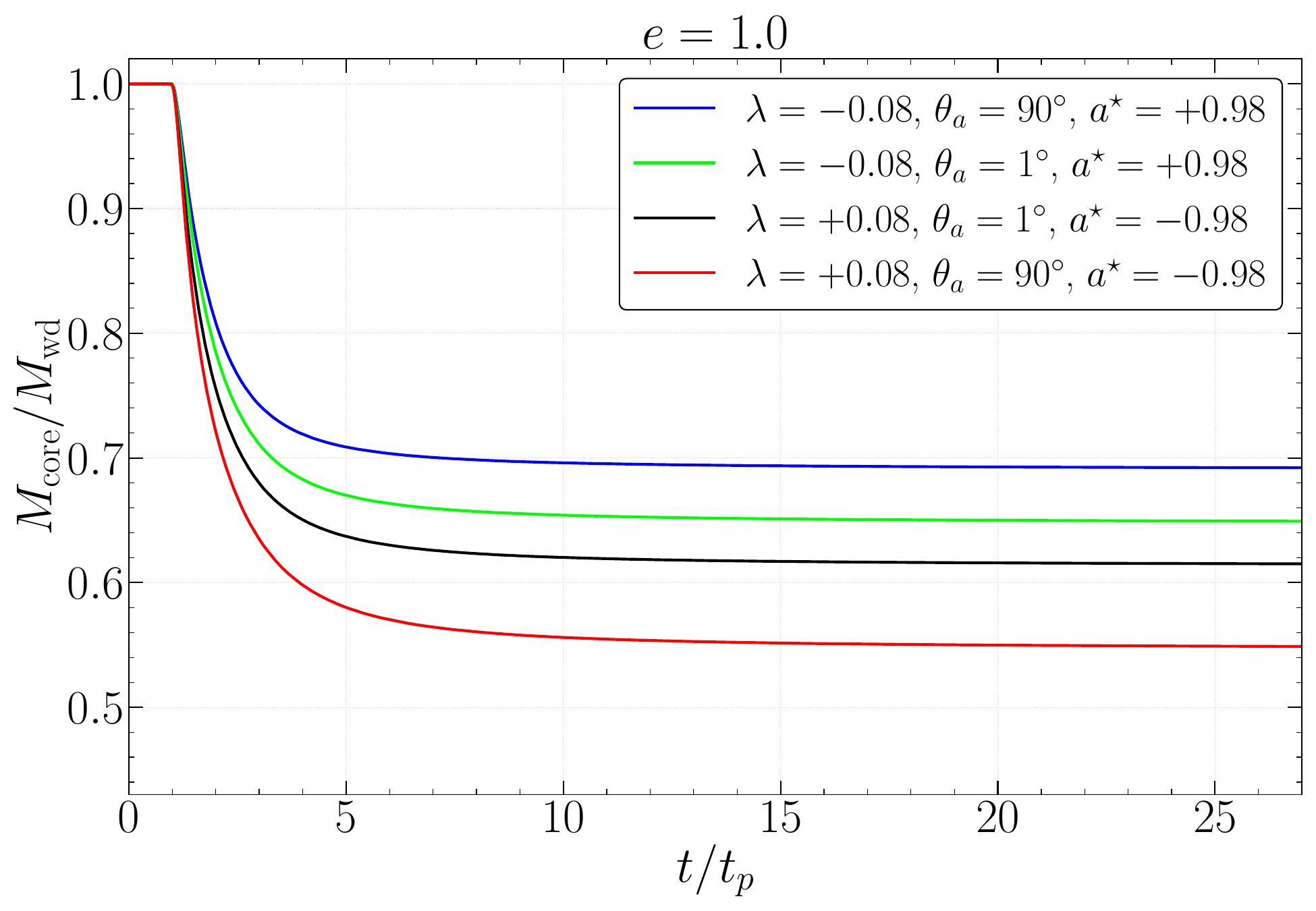}
\hfill
\includegraphics[width=.45\textwidth]{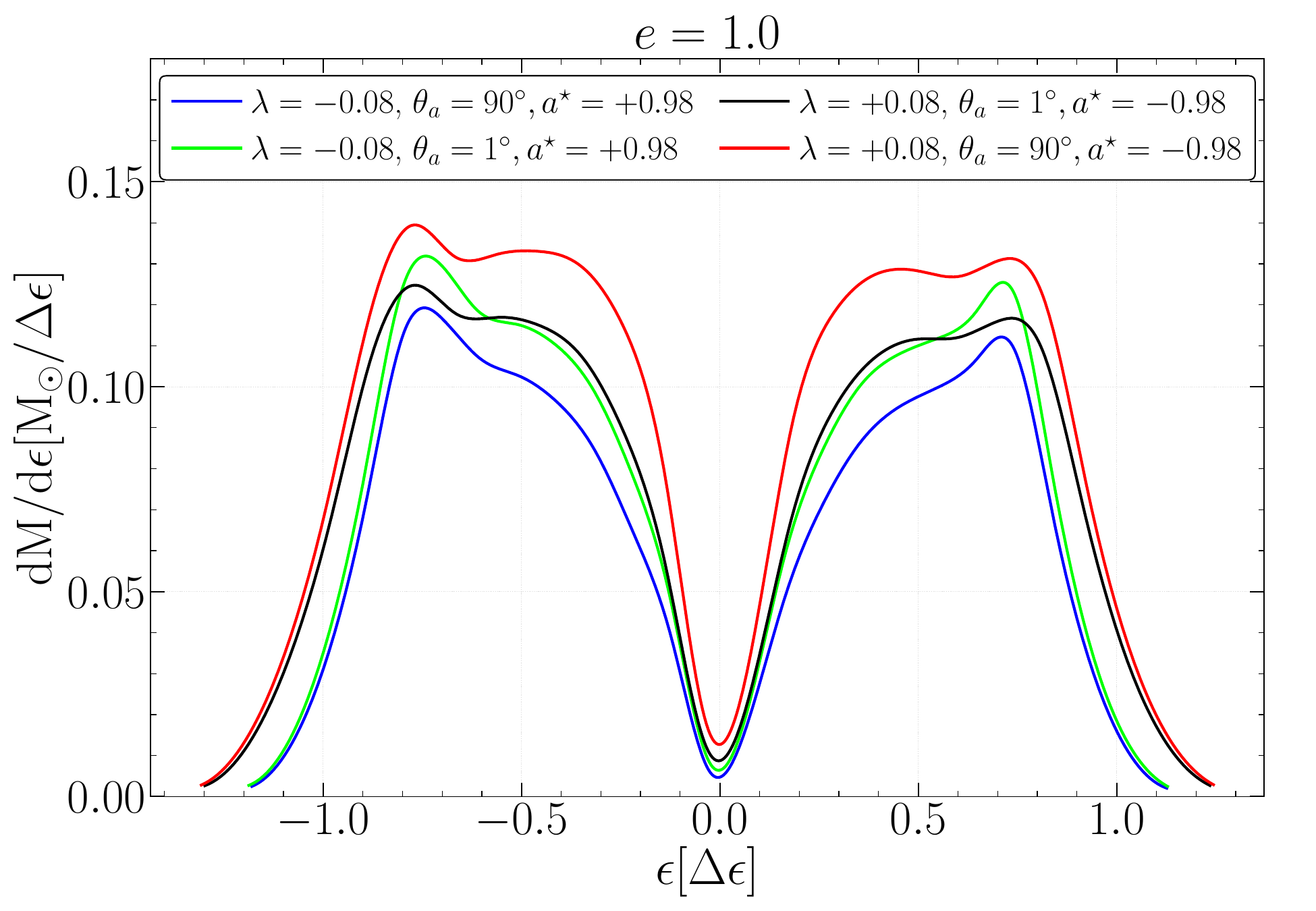}
\caption{Change in the variations of tidal debris due to WD spin for the specified values of breakup fraction $\lambda$, inclination angle $\theta_a$ and BH spin $a^{\star}$. \textbf{Left Panel:} The variation of core mass fraction,  $M_{\text{core}} / M_{\text{wd}}$ against normalized time. \textbf{Right Panel:} The variation of differential mass distribution $\text{dM}/\text{d}\epsilon$ with respect to normalized specific energy.}
    \label{fig-4}
\end{figure}
When spin is introduced to WD, the outcomes reflect similar trends as discussed earlier in \cite{2024arXiv240117031G}. 
Significant changes are primarily observed in cases where the spin of WD is either aligned or anti-aligned with its orbital angular momentum. As expected, mass disruption is enhanced in the prograde motion of the WD ($\lambda=+0.08$) and suppressed in the retrograde motion of the WD ($\lambda=-0.08$). The changes in the $\theta_a = 1^\circ$ off-equatorial orbits are observed solely due to the coupling effect between the WD spin and its orbital angular momentum, with little influence from the BH spin. However, in the equatorial plane, the coupling of the BH spin with the WD spin and the orbital angular momentum leads to distinct outcomes, as explained in \cite{2024arXiv240117031G}. 

Among all possible configurations, we find that in the equatorial plane, the case with WD spin $\lambda = +0.08$ and BH spin $a^\star = -0.98$ results in the maximum mass disruption, while the case with $\lambda = -0.08$ and BH spin $a^\star = +0.98$ results in the minimum mass disruption. The corresponding maximum and minimum mass disruptions in the $\theta_a = 1^\circ$ off-equatorial orbits fall between these two extremes; see the left panel of Figure \ref{fig-4}. In Table \ref{table-2}, we have displayed all the relative differences\footnote{Relative differences are computed with respect to the results corresponding to non-spinning WD.} in core mass fraction between spinning and non-spinning WDs for corresponding $a^\star$ and $\theta_a$. Based on these results, we calculate the debris differential mass distribution with their specific energies. The distribution profiles are generated from the post-disruption snapshot taken at $t \approx 0.21$ hr after the bound core mass has saturated and before the debris has accreted. In this case, the spread in the $\theta_a = 1^\circ$ and $\theta_a = 90^\circ$ orbits is not the same for both BH spins, unlike the non-spinning scenario. This indicates that the spread is primarily influenced by the stellar spin, rather than the spin of the BH. We observe a slightly wider (for $\lambda = +0.08$) or narrower (for $\lambda = -0.08$) spread in the $\text{dM/d}\epsilon$ distribution profiles due to the WD spin for the specified configurations, as shown in the right panel of Figure \ref{fig-4}. This variation occurs due to the small change observed in the core mass fraction in the presence of WD spin. 
\begin{table}[htbp]
    \centering
    \begin{tabular}{|c|c|c|c|c|}
        \hline
        \multirow{2}{*}{$a^{\star}$} & \multicolumn{2}{c|}{Relative Difference in $M_{\text{core}}/M_{\text{wd}}$} & \multicolumn{2}{c|}{Relative Difference in $M_{\text{core}}/M_{\text{wd}}$} \\ 
        & \multicolumn{2}{c|}{between $\lambda=0.00$ and $\lambda=+0.08$} & \multicolumn{2}{c|}{between $\lambda=0.00$ and $\lambda=-0.08$} \\ \cline{2-5}
        & $\theta_a = 1^{\circ}$ & $\theta_a = 90^{\circ}$ & $\theta_a = 1^{\circ}$ & $\theta_a = 90^{\circ}$ \\ \hline
        $+0.98$ & 2.99\% & 2.62\% & 2.09\% & 1.84\% \\ \hline
        $-0.98$ & 3.01\% & 3.70\% & 2.10\% & 2.47\% \\ \hline
        \end{tabular}
    \caption{Comparison of the relative difference in $M_{\text{core}}/M_{\text{wd}}$ for different $\lambda$ values, BH spins $a^{\star}$, and angular orientations $\theta_a$ in a parabolic orbit.}
    \label{table-2}
\end{table}

\subsection{Observables}\label{observables}
We now analyze the observable variations with the inclination angle, BH spin, and WD spin as a result of the tidal interaction. 
Mass fallback rates are calculated directly from the simulation without using the well-known ``frozen-in'' approximation. \red{This frozen-in approximation assumes that a star is fully disrupted by instantaneously losing its self-gravity at pericentre. Further, it allows one to approximate the orbits of the 
resultant debris particles by Keplarian ones and, using their corresponding orbital times, to estimate the fallback rate.
	However, the presence of a self-bound core can change these Keplerian orbits of the bound debris. To avoid such assumptions, we instead compute numerical derivatives directly from the accreted mass in the simulation.} To compute the accreted mass, 
we excise a $3r_t$ radius around the BH when the tidal tails and the self-gravitating core move away from the BH after the interaction. Once the bound part starts moving towards the BH and crosses 
the $3r_t$ radius, it is accreted by the BH. We note the mass accreted by the BH with time, and the numerical derivative of the data is used to compute the fallback rates. This methodology is used in e.g., \cite{2015ApJ...808L..11C, 2019ApJ...882L..26G, Miles, 2023JCAP...11..062G}. Here, we will focus on parabolic 
orbits as elliptic ones are more challenging due to debris intersection, multiple encounters, etc., and we hope to report on this in future. 

\begin{figure}[h]
\centering
\includegraphics[width=.45\textwidth]{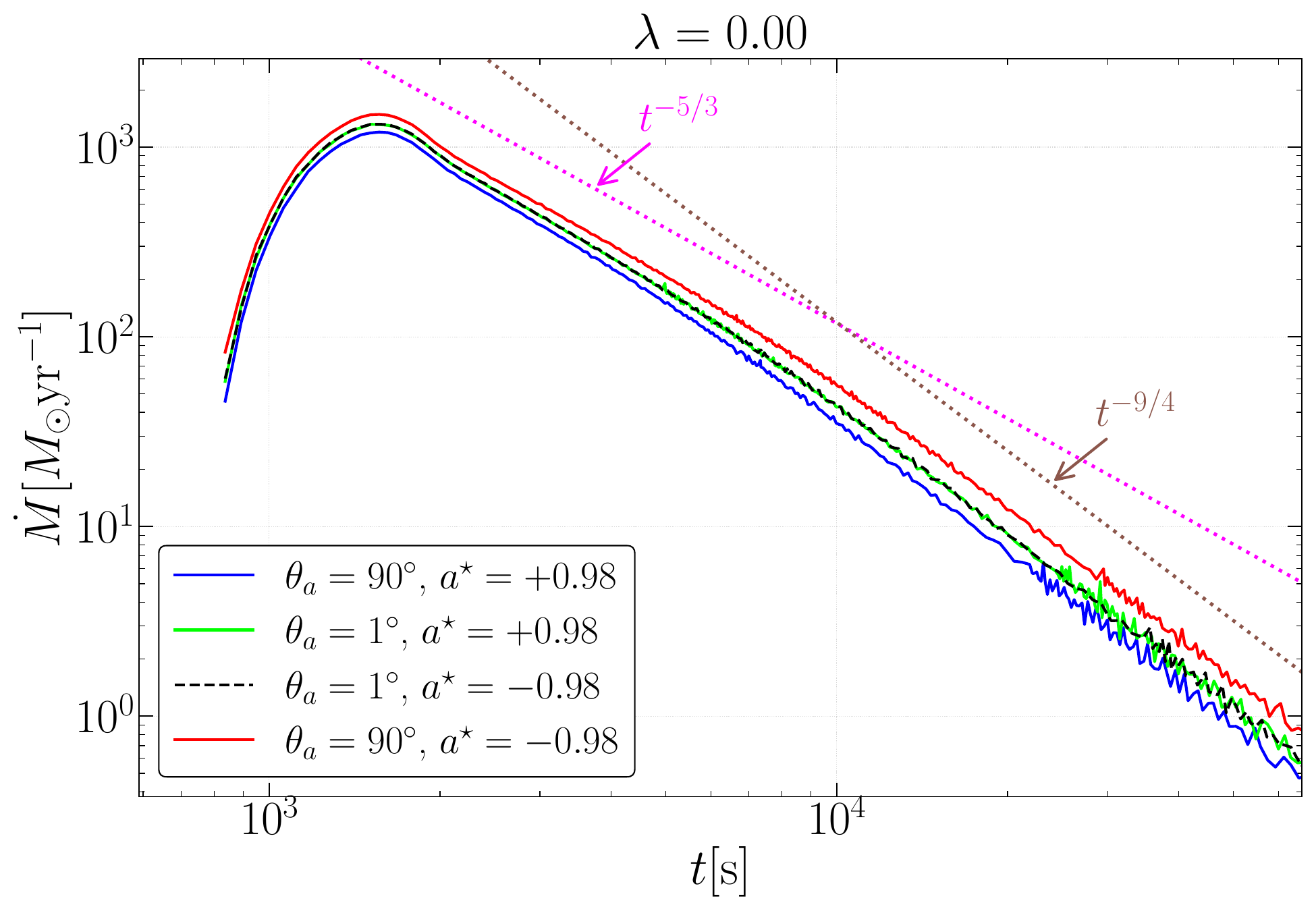}
\hfill
\includegraphics[width=.45\textwidth]{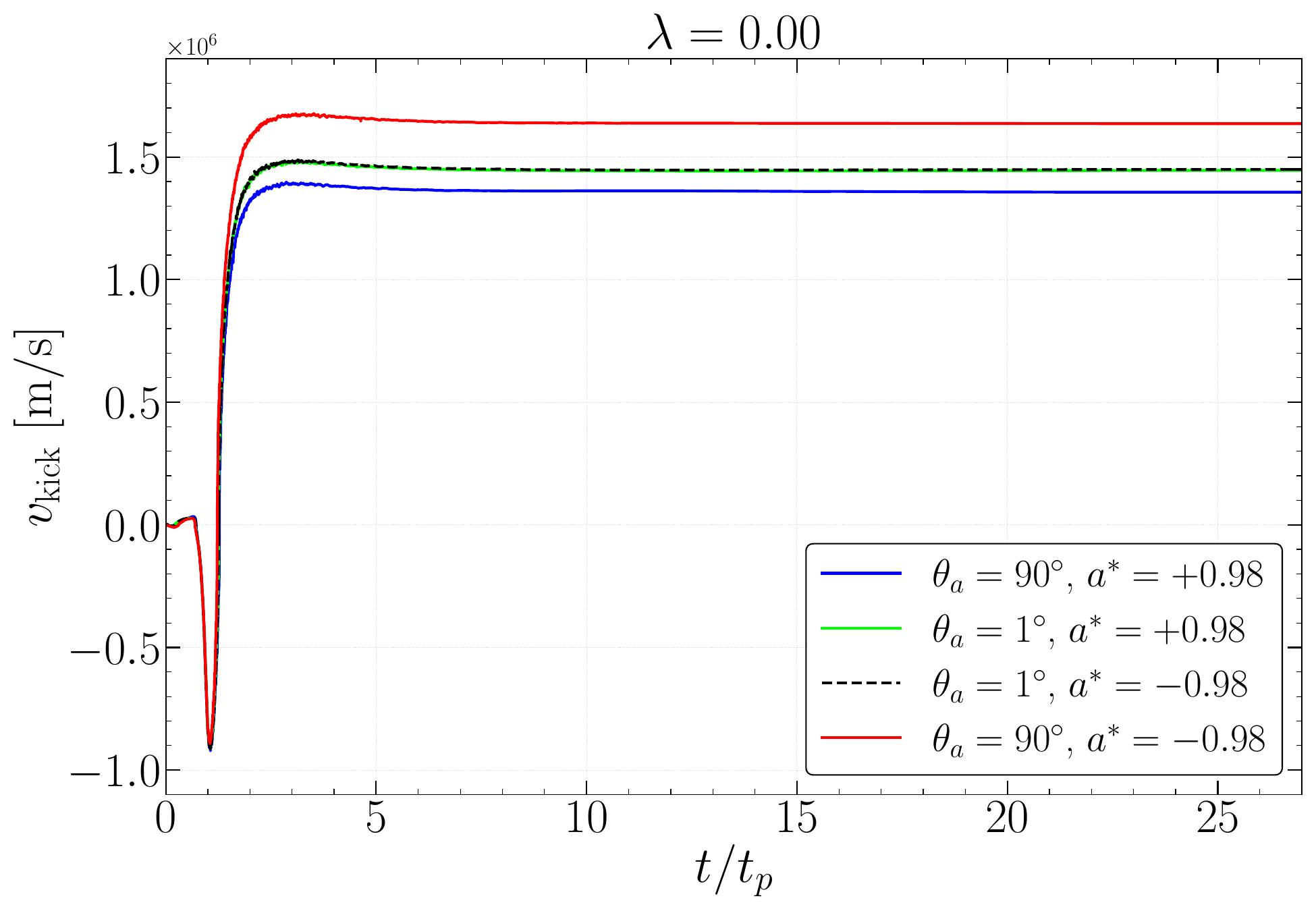}
 \caption{\textbf{Left Panel:} Fallback rate plotted against time in hours. \textbf{Right Panel: }Kick velocity of bound core plotted against normalized time.}
    \label{fig-5}
\end{figure}

In the left panel of Figure \ref{fig-5}, the fallback rate is shown to have a lower peak value for the motion in the $\theta_a=90^{\circ}$ plane compared to the $\theta_a=1^{\circ}$ off-equatorial orbit for BH spin of $a^\star=+0.98$ (and in the absence of spin of the WD). Again, the reverse behavior is observed for BH spin $a^\star=-0.98$. Overall, the dynamics suggest that the fallback rate is dominated by the retrograde motion ($a^\star=-0.98$ \& $\theta_a=90^{\circ}$), while the prograde motion ($a^\star=+0.98$ \& $\theta_a=90^{\circ}$) results in a lower fallback rate. 

The fallback rate for the $\theta_a=1^{\circ}$ off-equatorial orbit consistently falls between those for prograde and retrograde motions.  
\blue{We also note here that the fallback rate follows the usual trends known for
partial TDEs \cite{Miles},\cite{2024ApJ...967..167G}. Namely, the initial slope of the fallback rate scales as $t^{-5/3}$ 
and then settles to a $t^{-9/4}$ power law at later times. Note that this late time slope is independent of the initial orbital inclination as well
as the black hole spin.
}

By virtue of the conservation of linear momentum during the partial tidal disruption phase, the bound tail imparts a “kick” to the saturated bound core. Consequently, the bound core can gain sufficient energy to eject from the influence of the BH and become a hypervelocity star \citep{2014ApJ...782L..13M, 2015MNRAS.449..771G}, which increases its specific orbital energy and specific angular momentum. This is quantified by the kick velocity, defined as; $v_{\text{kick}}=\sqrt{\red{2}(\epsilon_{\text{core}}-\epsilon_{\text{in}})}$, where $\epsilon_{\text{core}}$ and $\epsilon_{\text{in}}$ denote the specific orbital energies of the bound core and the initial WD, respectively. \red{The quantity $\epsilon_{\rm in}$ gives the specific orbital energy of the entire WD at $r_0$, the point where it begins its orbital motion around the BH. As the WD approaches pericentre, it loses mass, and the remaining self-bound core acquires a higher specific orbital energy $\epsilon_{\text{core}}$, compared to its initial undisrupted state. Consequently, the core’s specific kinetic energy rises suddenly and then saturates at a higher value as the self-bound core attains a constant mass.} In the right panel of Figure \ref{fig-5}, we demonstrate the effect of inclination on the kick velocity imparted to the saturated WD core for the specified values of $a^{\star}$ and $\theta_a$. The relative gain in the velocity of the bound core due to the kick between prograde and retrograde motion is around $20.5 \%$ for $\theta_a = 90^{\circ}$. The bound core corresponding to $\theta_a = 1^{\circ}$ has an almost identical kick velocity across both BH spins.

\begin{figure}[h]
\centering
\includegraphics[width=.45\textwidth]{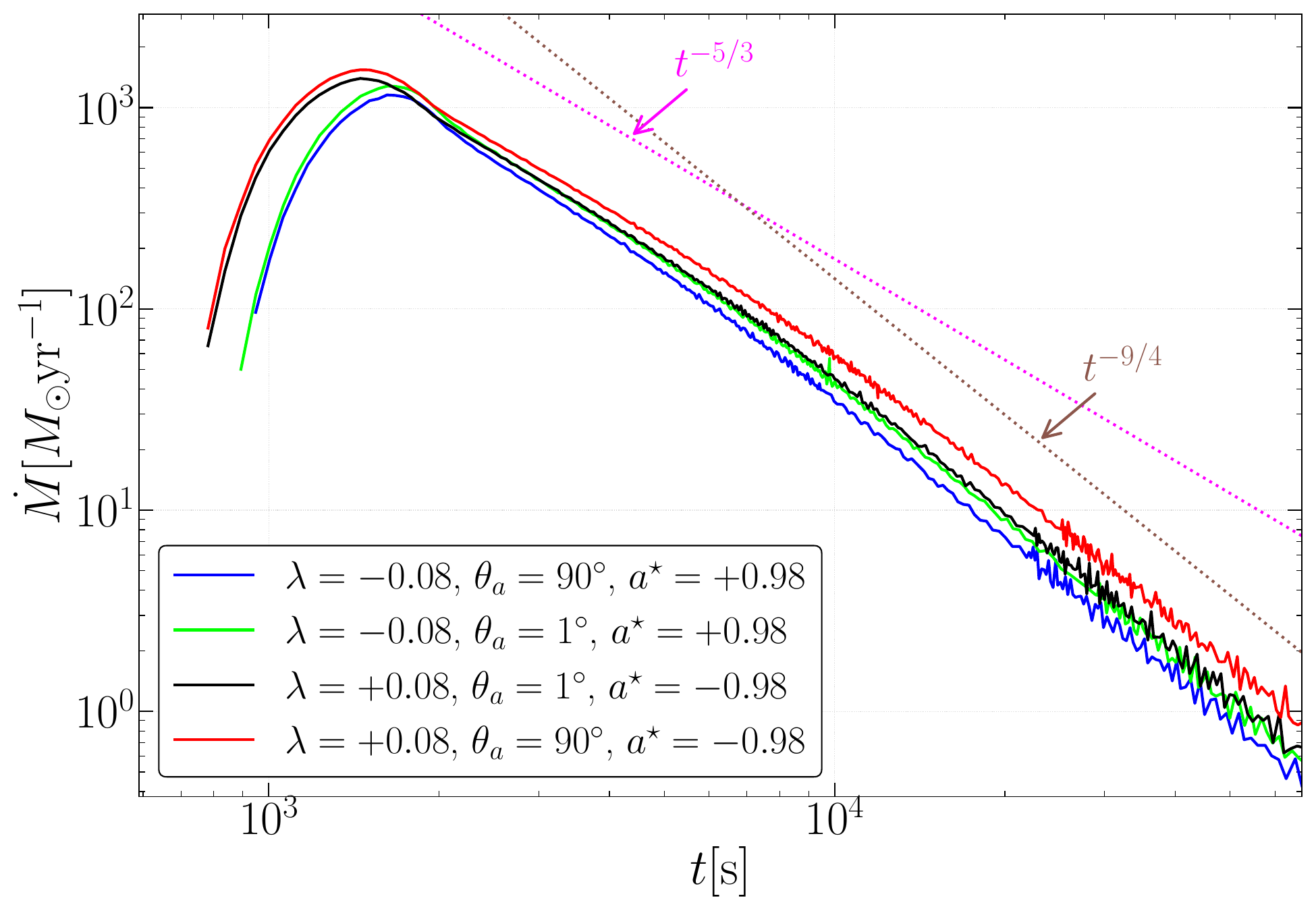}
\hfill
\includegraphics[width=.45\textwidth]{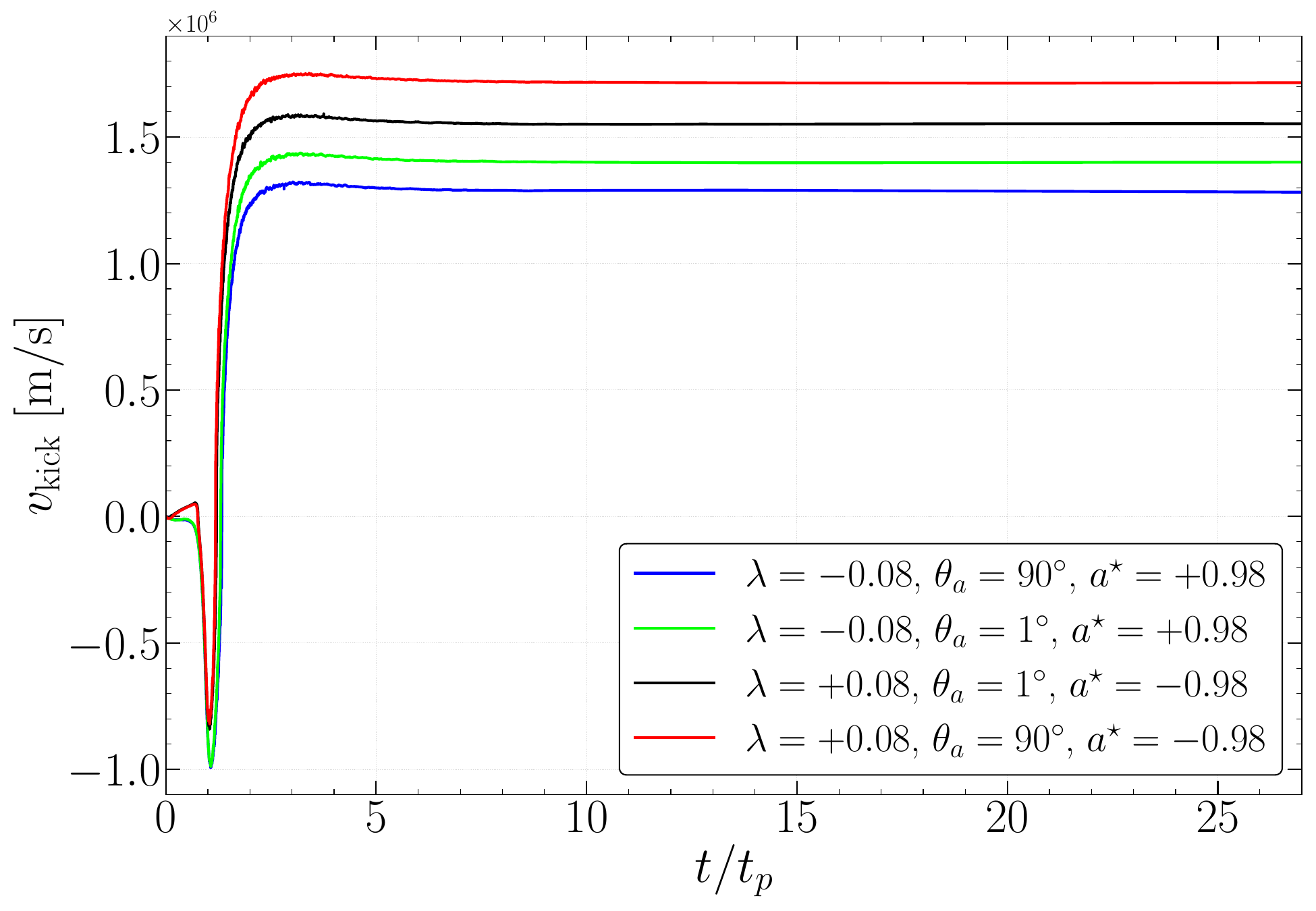}
\caption{Variation in the presence of WD spin. \textbf{Left Panel:} Fallback rate plotted against time in hours. \textbf{Right Panel: }Kick velocity of bound core plotted against normalized time.}
    \label{fig-6}
\end{figure}

In the case of a spinning WD, we observe similar effects on the mass fallback rate due to stellar spin, as seen in the case of mass disruption. These changes are influenced by the coupling between the WD spin and the orbital angular momentum, as displayed in the left panel of Figure \ref{fig-6}. The inclination effect remains the same as before. The maximum peak fallback rate is observed for the BH spin $a^{\star} = -0.98$ and $\theta_a = 90^\circ$, while the minimum occurs for $a^\star = +0.98$ and $\theta_a = 90^\circ$. 
However, we note that the WD spin, with values $\lambda = +0.08$ and $\lambda = -0.08$, leads to a slight increase and decrease in the peak fallback rate, respectively, 
for both $\theta_a = 1^\circ$ and $\theta_a = 90^\circ$. This occurs due to the fact that tidal torque finally aligns the stellar spin in the prograde direction with 
respect to the orbital motion of the WD. In the case of initial retrograde stellar spin, the star is first spun down and then spun up while reaching the pericenter. 
On the other hand, for an initial prograde stellar spin, the tidal torque enhances the same so that more matter can be stripped off the WD as it reaches the pericenter. This anti-alignment also results in a small difference between the fallback times of the most bound debris onto the BH. 
\blue{As in the case of non-spinning WDs, the slope of the fallback rate settles to a $t^{-9/4}$ power law, starting from a $t^{-5/3}$ one. As in
that case, this is independent of the initial inclination of the orbit, as well as BH or stellar spin. We will discuss this further in section \ref{comparison}
}
We also note that in this case, the maximum relative change in the kick velocity for the WD bound core on $\theta_a=90^\circ$ plane between the given $\lambda$ and $a^\star$ becomes $33.86 \%$, as found in the right panel of Figure \ref{fig-6}.

Overall, it is useful to summarise the effects of the moderate stellar spin that we have used here, following the discussion above.
First, for non-rotating stars,
the left panel of Figure \ref{fig-5} shows that the time of arrival of the maximally bound debris is similar, independent of BH spin. This 
degeneracy is broken with stellar spin, as indicated by the left panel of Figure \ref{fig-6}. Also, the kick velocity corresponding to the $\theta_a = 1^\circ$
orbit is independent of the BH spin for a non-spinning star, as suggested by the right panel of Figure \ref{fig-5}, while stellar spin again breaks this 
degeneracy (right panel of Figure \ref{fig-6}). For more rapid stellar rotations, these interesting physical effects might be magnified, and it will 
be useful to study this further.

\subsection{Gravitational Wave Emission}

TDEs are promising sources of gravitational radiation. When a WD passes close to the pericenter of its orbit around a BH, tidal disruption generates a peak in the GW amplitude. This burst-like behavior occurs because the stellar debris becomes too dispersed after pericenter passage, suppressing strong GW emission beyond $r_p$. The emission can be characterized by an amplitude $h$ and a duration $t \sim 1/f$, where $f$ is the frequency of the GW.

We computed the GW emission using the quadrupole approximation, following the method outlined in \cite{RR}. 
\blue{This approach provides a leading-order characterization of gravitational wave emission using the weak-field approximation, which is expected to be sufficiently accurate for our current analysis. A detailed post-Newtonian treatment
will be explored in future studies.}
The GW polarization amplitudes, $h_+(t)$ and $h_\times(t)$, as well as the total amplitude, $|h(t)| = \sqrt{|h_+(t)|^2 + |h_\times(t)|^2}$, were evaluated over time for various tidal disruption scenarios. Our system setup and assumptions follow \cite{2022MNRAS.510..992T}, with the pericenter of the WD's orbit along the $y$-axis and the system observed face-on, which maximizes the GW signal. For inclined orbits, the system was rotated into the $x$-$y$ plane before calculating the GW emission using the expressions from \cite{2007gwte.book.....M}.

\begin{figure}[h]
    \centering
    \includegraphics[width=1.0\linewidth]{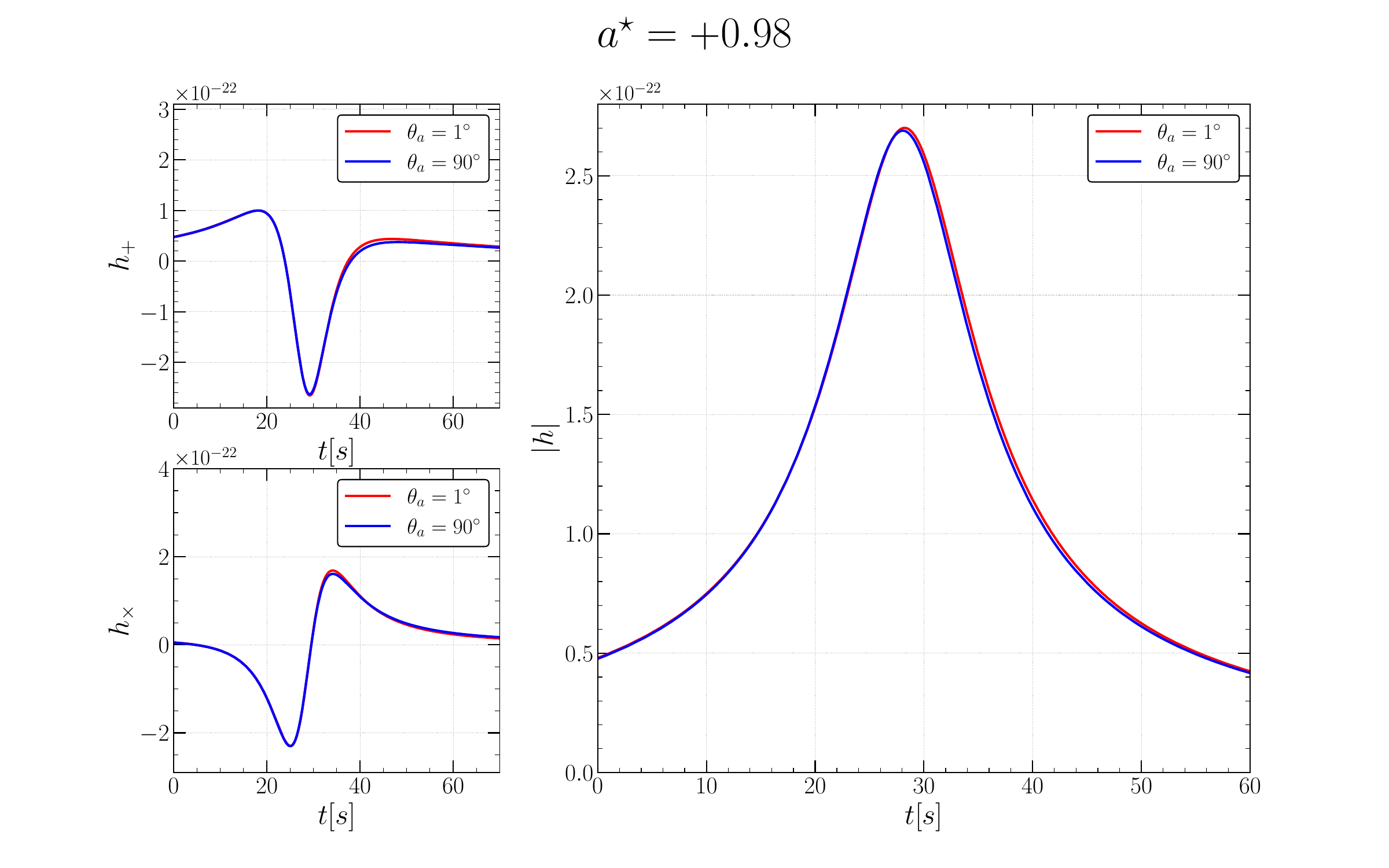}
    \caption{GW polarisation amplitudes by a WD-IMBH system on a parabolic orbit at a distance of $d$ = 20 Mpc, plotted with respect to the time (in seconds). \textbf{Top Left Panel:} $h_+$ polarization seen face-on \textbf{Bottom Left Panel:} $h_\times$ polarization seen face-on  \textbf{Right Panel:} Root-square-sum amplitude.}
    \label{fig-7}
\end{figure}

Figure \ref{fig-7} illustrates the GW polarization amplitudes $h_+$ and $h_\times$, along with the strain magnitude $|h|$ for a parabolic orbit around a BH with spin parameter $a^{\star} = +0.98$, plotted against time (in seconds). The relative amplitude difference $|h|$ between the equatorial and off-equatorial orbits is about 0.34\%. The strain in the inclined orbit, $\theta_a = 1^\circ$ is slightly higher than in the $\theta_a = 90^\circ$ plane, consistent with the findings of \cite{2022MNRAS.510..992T}. In our case, the penetration factor $\beta$ is not large enough to produce a significant change. A slight shift in the peak is also observed, which is attributed to a small variation in the passage time to the pericenter. For the BH spin $a^{\star} = -0.98$, this pattern reverses, with the amplitude in the $\theta_a = 90^\circ$ plane becoming larger. Notably, the relative amplitude difference increases to approximately 0.6\%, slightly higher than in the $a^{\star} = +0.98$ case.  We also verified that rotation does not introduce significant changes in the GW polarisation amplitude. 

With our choice of parameters, the GW amplitude is $\sim 1.83 \times 10^{-22}$. The signal duration is about $\sim 40$ seconds, corresponding to a characteristic frequency of 
$\sim 2.4 \times 10^{-2}$ Hz (see Equations 12 and 13 of \cite{2022MNRAS.510..992T}).
At a source-to-observer distance of $d = 20$~Mpc (chosen following \cite{2022MNRAS.510..992T}), 
these values are consistent with what we find
from our numerical simulations using the quadrupole approximation, highlighting the potential for detecting such signals in the context of multi-messenger astronomy. \red{However, the small variations we find are unlikely to be directly distinguishable with current detector sensitivities. Even if future instruments achieve the required precision, their observational imprint would remain challenging to isolate due to strong parameter degeneracies—for instance, a waveform generated by a given black hole mass and orbital inclination can closely resemble that of a system with slightly different parameters. Thus, while our results demonstrate the presence of inclination-dependent effects, their direct observational imprint is expected to be largely obscured by degeneracies in the parameter space. Nonetheless, combining gravitational-wave measurements with electromagnetic counterparts could help to break these degeneracies and provide tighter constraints on system parameters in future multi-messenger observations \cite{MM}.}

\subsection{Comparison with previous literature}\label{comparison}

\blue{As we have mentioned in section \ref{back}, full TDEs of polytropic solar mass stars by $10^6~M_{\odot}$ 
Kerr SMBHs have been considered by 
\cite{Hayasaki2016}, \cite{LiptaiPrice}. On the other hand, full TDEs of initially non-rotating solar mass WDs from $10^3~M_{\odot}$ Kerr
IMBHs have been discussed by \cite{Haas}, and it will be useful to compare and contrast our results on partial TDEs with these. 
Let us briefly recall the relevant results of \cite{Haas}. Firstly, these authors find that two extreme cases occur when WD spin is aligned 
parallel or opposite to the black hole spin. Due to the closeness of the encounter, the effect of black hole spin 
is dramatic. Namely, in the former, very little material is accreted in the first few seconds following disruption, while for the latter, 
almost all the stellar debris is accreted into the BH by this time. For the other cases considered, the amount of matter accreted in the 
first few seconds falls within these two extremes. Secondly, \cite{Haas} finds that the slope of the late time slope of 
the mass fallback rate is proportional to $t^{-5/3}$, which is, as discussed in the introduction the prediction from a Newtonian
approach \cite{Rees,Phinney}.  
}

\blue{To compare the above results with those obtained here, we first note that since we have considered partial disruptions,
i.e., milder encounters of WDs with Kerr IMBHs, the effects of spin are significant, but less dramatic than the ones found
in \cite{Haas}. Namely, for initially non-rotating WDs, the extrema of mass stripping occur when the spin of the BH
is aligned parallel/opposite to the orbital angular momentum of the WD, with the maximum mass being stripped for
opposite orientation (see Fig. \ref{fig-1}), consistent with the observations of \cite{Haas}. In our case, inclusion of stellar
spin has interesting physical effects, a scenario that was not considered in that work. Further, we find that in our case, 
the peak of the fallback rate happens at a time which is more than two orders of magnitude later than, and that its value is
more than an order of magnitude less than that shown in Figure 8 of \cite{Haas}. Further, the late
time slope of the fallback rate is proportional to $t^{-9/4}$ as opposed to the $t^{-5/3}$ power law found there. The
reason for the difference in the time of occurrence and the magnitude of peak value of the fallback rate 
is the milder nature of the encounter that we have considered here. Indeed, \cite{Haas} has commented upon
the prompt accretion seen in their scenario, and attributed it to effects of the BH spin. In our case, due to
the large value of the pericenter distance, such effects are much weaker.
}

\blue{
As far as the late time slope is concerned, the power law obtained here is similar to the one
predicted in \cite{CN} for partial TDEs by SMBHs. Now, as we have mentioned in section \ref{back}, the work
of \cite{2024ApJ...967..167G} showed that partial TDEs from IMBHs might indeed show different power laws.
In this case, the $t^{-9/4}$ power law is however an artefact of the choice of parameters, and is in reasonable
agreement with the prediction of Equation 5 of \cite{2024ApJ...967..167G}, although that paper dealt with
$10^3M_{\odot}$ IMBHs.\footnote{We point out a typographical error in Equation 5 of \cite{2024ApJ...967..167G}.
The third term in the numerator of the right hand side should come with a negative sign.} Note that the late time 
slope of the fallback rate is independent of the initial orbital inclination, BH spin, as well as stellar spin, 
as is evident from Fig. \ref{fig-5} and \ref{fig-6}. This is similar to what was reported in \cite{Haas}, where the
authors found that this slope was always $t^{-5/3}$, independent of the orbital inclination and BH spin. As pointed
out in  \cite{2024ApJ...967..167G}, this slope can be different for different values of the impact parameter, for equatorial
orbits. Since the slope is independent of orbital inclination, the above discussion points to the fact that  
a different choice of parameters than those considered here would have yielded a different slope, which would always
be steeper than the $t^{-5/3}$ behaviour found in \cite{Haas} for full disruptions. 
}

\section{Discussions} \label{sec:conclusion}

As we have mentioned in the beginning, the motivation for this paper is twofold. Firstly, we consider TDEs involving 
the less-studied Kerr IMBHs and secondly, we consider such TDEs in off-equatorial orbits for spinning WDs. Analytical studies of TDEs in 
off-equatorial orbits of the Kerr BH is fast becoming popular \cite{StockingerShibata,Shibata2}, and here we have presented a 
comprehensive numerical study of this important topic, and quantified observable signatures of partial TDEs and uncovered
interesting physics in the presence of stellar spin. Although we
conclude that current observations may not be able to pinpoint off-equatorial TDEs, our analysis nonetheless
shows important physics of the coupling between BH and stellar spin even with moderate stellar rotation.

Our findings in this paper emphasize the complex interplay between spin, 
orbital inclination, and orbital motion in determining the nature of debris distribution, mass fallback rates,  and the ejection of the bound 
core during TDEs.
Our main observation here is that as far as WD mass loss, debris mass distribution with specific energy, kick velocity to the self-bound core, 
and fallback of debris onto the BH are concerned, off-equatorial results lie somewhere in between extreme cases of equatorial ones.
We therefore conclude that as far as observations are concerned, TDEs in off-equatorial orbits 
are degenerate with those in equatorial orbits with different BH spins.

\begin{figure}[h]
\centering
\includegraphics[width=.45\textwidth]{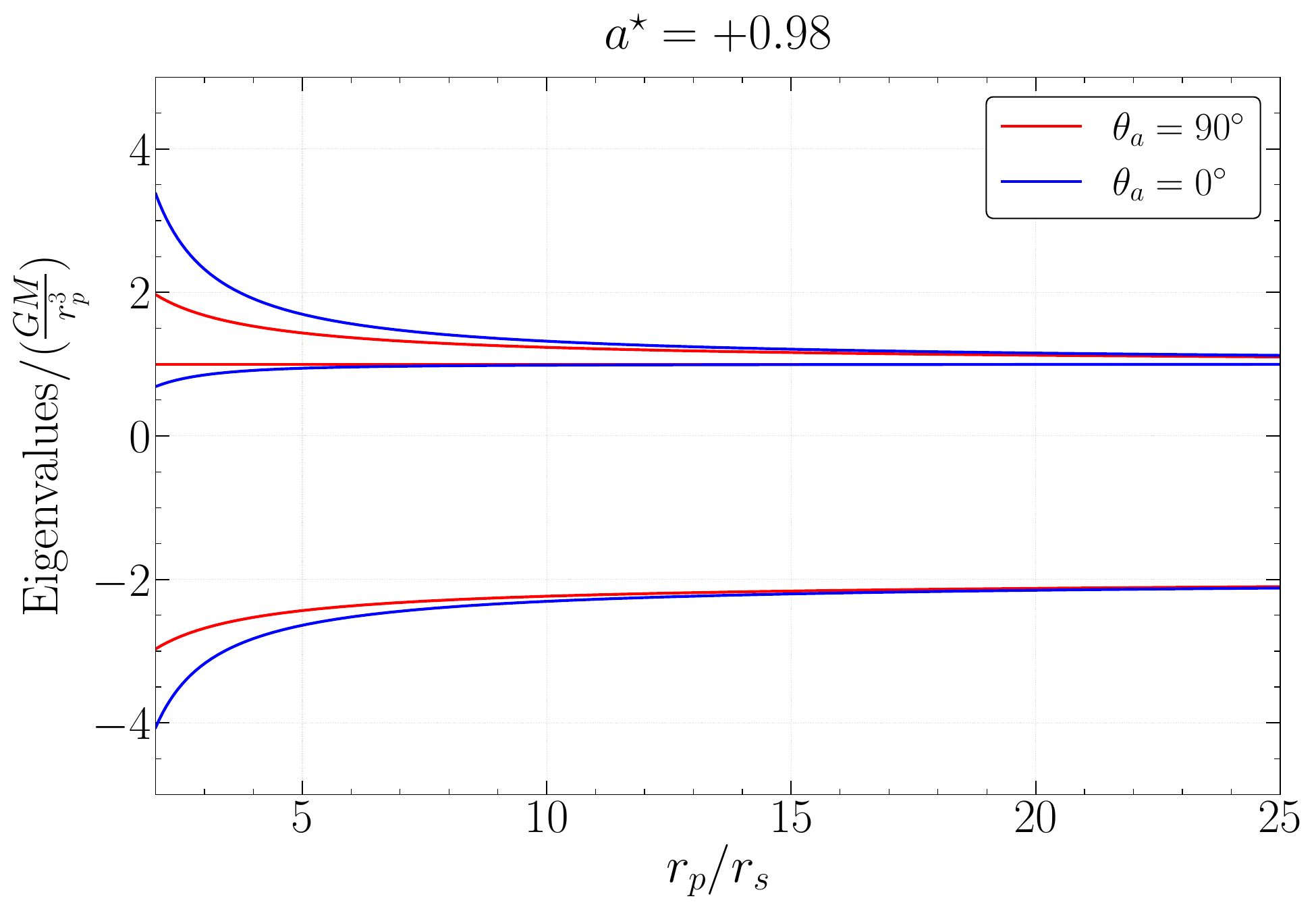}
\hfill
\includegraphics[width=.45\textwidth]{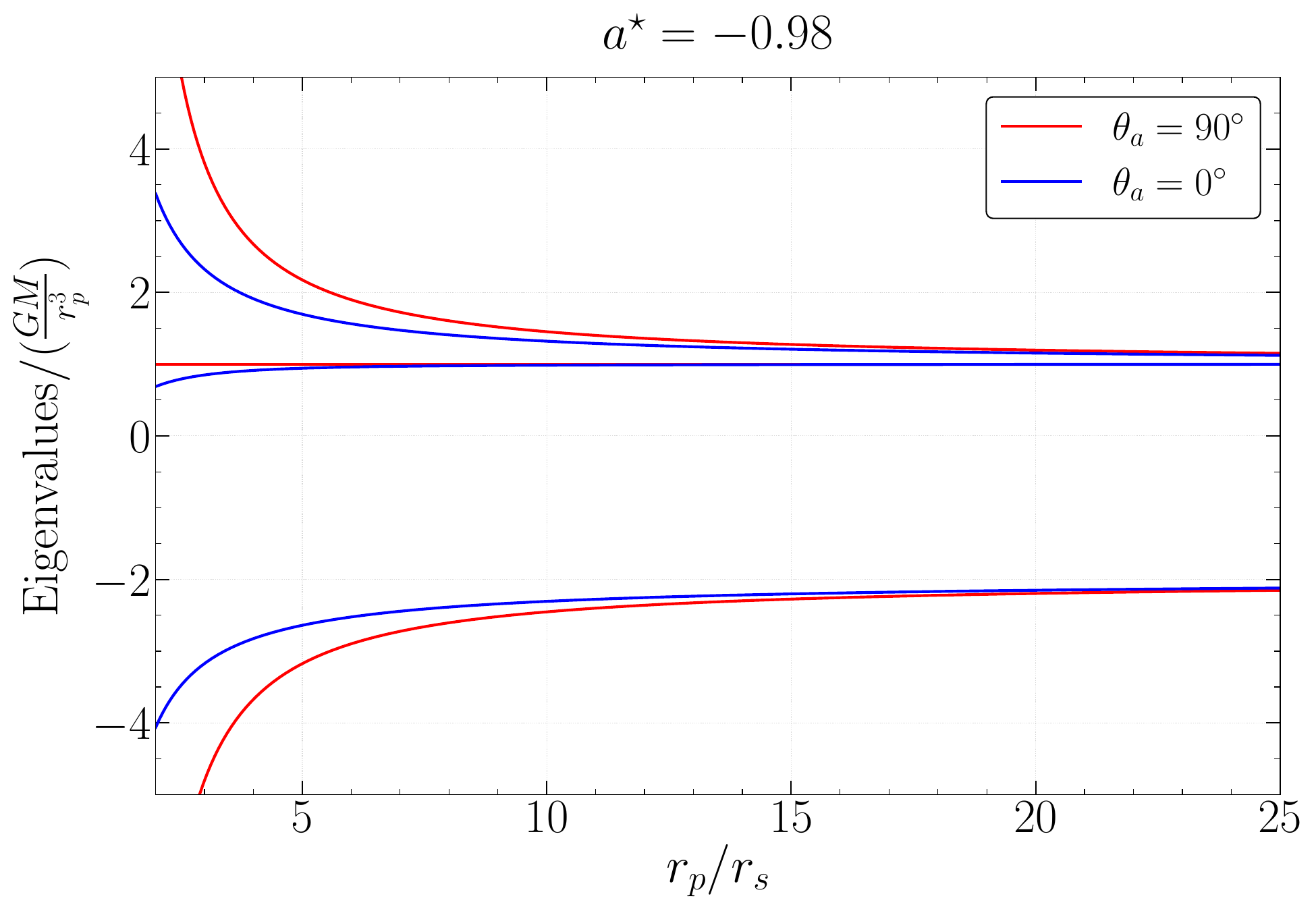}
\caption{Eigenvalues of the tidal stress tensor in Fermi normal coordinates \textbf{Left Panel:} BH spin $a^\star=+0.98$. \textbf{Right Panel: }BH spin $a^\star=-0.98$.}
    \label{stress}
\end{figure}

We also recall that a useful measure of the strength of tidal interactions is the tidal tensor. This can be understood as the difference between
the acceleration (force per unit rest mass) experienced by two neighboring particles of the stellar fluid and is conveniently defined by the 
spatial gradient of the acceleration. In the Newtonian case, the tidal tensor is defined by the
spatial derivatives, $C_{ij} = -\frac{\partial^2\Phi}{\partial x^i\partial x^j}$,
with $\Phi$ being the Newtonian potential. In the relativistic regime, the tidal potential is more challenging to compute; for the
Schwarzschild BH, it was obtained in \cite{2015MNRAS.449..771G}. Popularly in the literature, one resorts to the 
Fermi normal coordinates \citep{ManasseMisner} which gives a frame of reference locally flat along the geodesic trajectory in 
which the center of mass of the star moves. In these coordinates, one can obtain a Newtonian potential as a perturbation series around the 
trajectory of the center of mass in powers of the distance from the same. For the Kerr BH, the tidal tensor in the Fermi normal frame
expressed back in the standard Boyer-Lindquist coordinates in which the Kerr metric is written 
was computed in \cite{Marck}. 
To connect analytical predictions to our results obtained here, we show in Figure \ref{stress} the eigenvalues of the tidal 
stress tensor in units of $GM/r_p^3$ (with $G=c=1$) computed  
in the Fermi normal frame that co-rotates with the star (the so-called ``tilde'' frame in the literature), 
following \cite{Marck, StockingerShibata}. 
These are shown here as a function of the non-dimensional quantity $r_p/r_s$, and tend
to their well known Newtonian values $(1,1,-2)$ in the limit of large $r_p$. 
From Figure \ref{stress}, we see that our numerical results in section \ref{sec:results} are at par with those indicated here. 
Namely, for $\theta = 90^{\circ}$, the magnitudes of the eigenvalues of the tidal stress tensors
are smaller for $a^\star=+0.98$ compared to $a^\star=-0.98$. 
Also, the stress eigenvalues for $\theta_a=0^{\circ}$ are independent of the BH spin. 
In our numerical computations, this ``degeneracy'' was broken upon introduction of stellar spin. Also, for this inclination, the 
eigenvalues that maximally deviate from their Newtonian values lie in between
the stresses for $\theta_a=90^{\circ}$, $a^\star=\pm 0.98$. These translate to intermediate effects of inclined orbits that 
we have seen in this paper.


It is further useful to observe that the eigenvalues of the stress tensor have dimensions of the inverse square of time, so that when 
normalised by $GM/r_p^3$, these are dimensionless and
similar for BHs of all masses. Hence, the nature of the graphs in Figure \ref{stress} will remain unchanged even for SMBHs. In light of the discussion in
the above, this would indicate that the numerical results presented in this paper should generalise to TDEs involving 
such BHs as well, but further numerical studies
incorporating general relativistic effects in SPH are essential to reach a firm conclusion. 

We briefly discuss two other important points here. First, our study is of importance in the context of hypervelocity stars (HVSs),
which are stellar objects with velocities exceeding the escape velocity of their host galaxy \cite{1988Natur.331..687H, Hirsch:2005pb, Marchetti:2019, Marchetti:2021}. 
The first observational discovery of a HVS with a radial velocity of $709$ km/s in the Galactic rest frame was reported by \cite{2005ApJ...622L..33B}, 
with subsequent discoveries of numerous HVSs, as reviewed in \cite{2015MNRAS.452.4297B}. More recently, \cite{2020MNRAS.491.2465K} 
reported the discovery of S5-HVS1, a star with a total velocity of $1755 \pm 50$ km/s. In our case, the right panels of Figure \ref{fig-5} and Figure \ref{fig-6}
indicate similar kick velocities which can potentially result in detectable HVSs. However, whether such a star has been created due to an off-equatorial
interaction with an IMBH is difficult to predict. Secondly, we find a small change in the GW signal for off-equatorial orbits as compared to 
equatorial ones, in Figure \ref{fig-7}. In addition to the one reported here, 
we computed the GW signal with $\beta = 1.8$ and $r_p = 10.02 r_g$ with BH spin $a^{\star}=-0.98$, assuming that the Newtonian approximation to the stellar fluid
dynamics is still valid. Here we get a relative change of $2.6\%$ between equatorial and off-equatorial orbits in the GW amplitude. 
Clearly, deeper encounters increase the difference between the GW amplitudes, in lines with what has been reported 
in \cite{2022MNRAS.510..992T}. 

Finally, we mention some caveats that require further analysis. The work of \cite{GaftonRosswog} finds little dependence 
of the mass fall back rates on BH spin for TDEs involving SMBHs. Our results on the other hand are more in line with
\cite{Jankovic} (which uses a similar numerical scheme as ours), and we also seem to get consistency with those reported in \cite{Kesden1}. 
As suggested in \cite{Jankovic}, this might be due to the different implementation of stellar hydrodynamics in \cite{GaftonRosswog}. 
With the small impact factor considered here and the fact that the star is far from the strong gravity regime, we believe that
our results are robust, but further studies are certainly required to understand the situation more completely.

\section*{Acknowledgments}

We sincerely thank our referee for insightful comments and criticisms that helped improve a draft version of this paper. 
We acknowledge the support and resources provided by PARAM Sanganak under the National Supercomputing Mission, Government of India, at the Indian Institute of Technology Kanpur. The work of DG is supported by grant number 09/092(1025)/2019-EMR-I from the Council of Scientific and Industrial Research (CSIR). The work of TS is supported in part by the USV Chair Professor position at IIT Kanpur, India. The work of AM is supported by Prime Minister's Research Fellowship by Ministry of Education, Govt. of India. Our thanks to Sabyasachi Chakraborty for letting us use his cluster where some of the numerical computations were performed and to Atanu Samanta for related help. 

\appendix
\section{Final relaxed configuration of SPH WD}\label{app:A}
\purple{We obtain the complete $M$–$R$ mapping from SPH by initializing a series of SPH simulations corresponding to different central parameters $x(0)$, each defined by its respective theoretical density profile input. For each case, the final mass and radius of the relaxed WD were measured at the end of the stretch-mapping process, where the internal theoretical density profile $x(r)$ and other equilibrium conditions were satisfied.}

\begin{figure}[h]
   	\centering
   	\includegraphics[width=.45\textwidth]{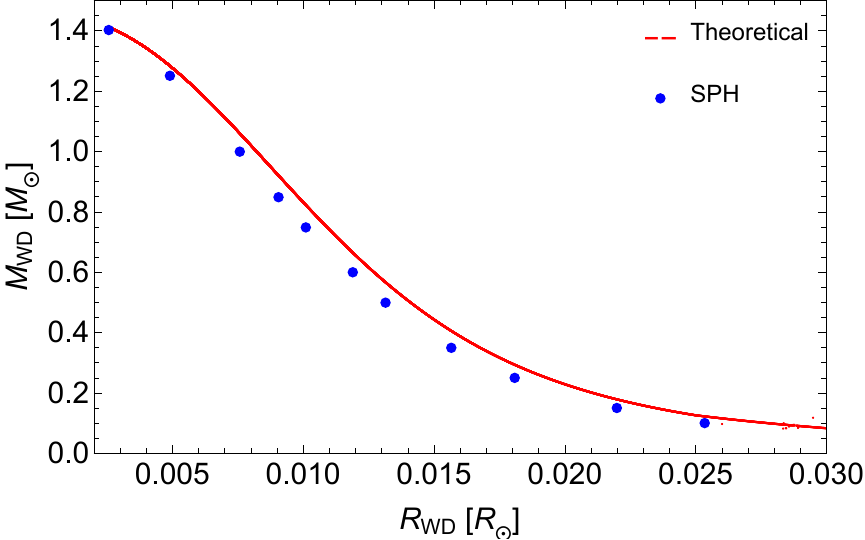}
   	\caption{Mass-radius relation of Carbon-Oxygen WDs. The red line is plotted using the numerical solutions of $x(r)$ and $m(r)$ for different $x(0)$ values.}
   	\label{mass-radius}
   \end{figure}

\purple{In Figure~\ref{mass-radius}, we present the resulting $(M_{\mathrm{WD}}^{\mathrm{SPH}}, R_{\mathrm{WD}}^{\mathrm{SPH}})$ values alongside the theoretical $(M, R)$ pairs. A total of 11 SPH configurations were generated, filling most of the $M$–$R$ curve.
The relative difference in the values of mass and radius between the SPH and theoretical results are given in Table \ref{tab:mass-radius}.
}

\begin{table}[h]
	\centering
	\setlength{\tabcolsep}{10pt} 
	\renewcommand{\arraystretch}{1.3} 
	\begin{tabular}{c|c|c|c|c|c|c}
		\hline
		\hline
		S.No. & $M$ ($M_{\odot}$) & $R$ ($R_{\odot}$) & $M_{\mathrm{WD}}^{\mathrm{SPH}}$ ($M_{\odot}$) & $R_{\mathrm{WD}}^{\mathrm{SPH}}$ ($R_{\odot}$) & $\Delta M / M$ & $\Delta R / R$ \\
		\hline
		1 & 0.10005 & 0.027694 & 0.099714 & 0.025347 & 0.0033583 & 0.084748 \\
		2 & 0.150018 & 0.023439 & 0.149695 & 0.021989 & 0.002153 & 0.061863 \\
		3 & 0.250003 & 0.019296 & 0.249479 & 0.018085 & 0.002096 & 0.06276  \\
		4 & 0.350034 & 0.016748 & 0.349045 & 0.015662 & 0.002827 & 0.06485 \\
		5 & 0.500031 & 0.014091 & 0.498603 & 0.013149 & 0.002855 & 0.06686 \\
		6 & 0.600018 & 0.012699 & 0.599239 & 0.011898 & 0.001298 & 0.06308 \\
		7 & 0.750065 & 0.010897 & 0.747858 & 0.010103 & 0.002943 & 0.07286 \\
		8 & 0.850017 & 0.009801 & 0.847951 & 0.00906 & 0.002431 & 0.07561 \\
		9 & 1.00001 & 0.008212 & 0.998865 & 0.00758 & 0.001145 & 0.07693 \\
		10 & 1.25000 & 0.005355 & 1.2502 & 0.004909 & 0.00016 & 0.08328 \\
		11 & 1.40000 & 0.002857 & 1.40177 & 0.002571 & 0.00126 & 0.1001 \\
		\hline
	\end{tabular}
	\caption{Comparison of SPH and theoretical white dwarf mass--radius pairs.}
	\label{tab:mass-radius}
\end{table}

\begin{figure}[h]
	\centering
	\includegraphics[width=.45\textwidth]{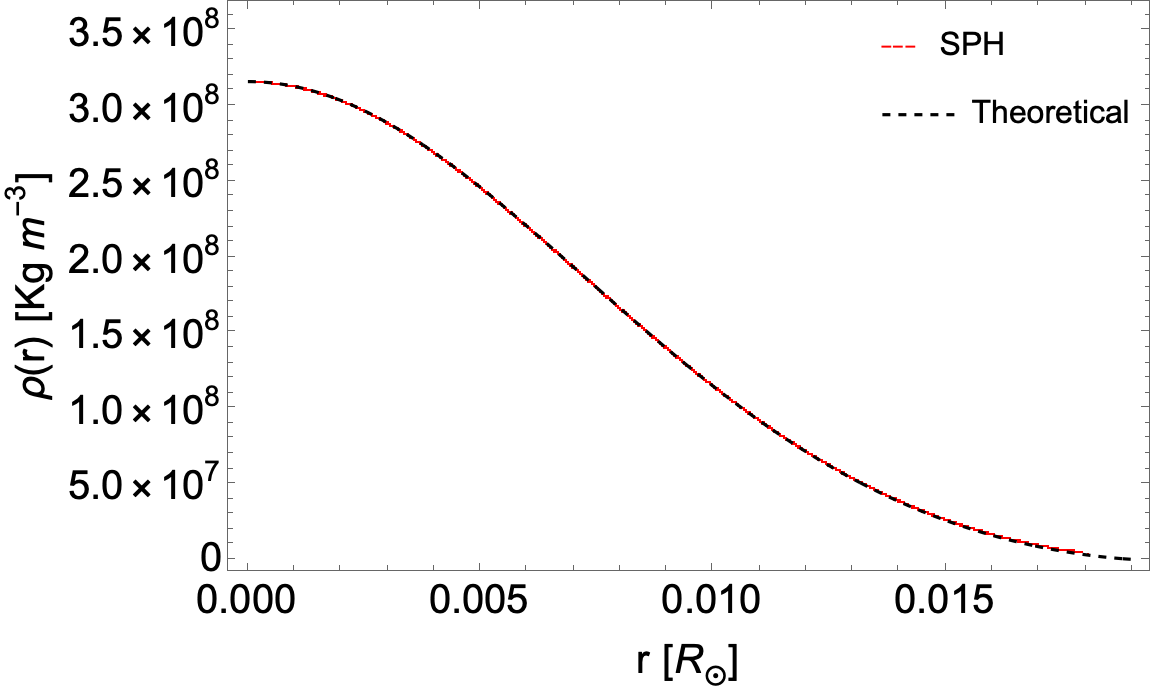}
	\hfill
	\includegraphics[width=.45\textwidth]{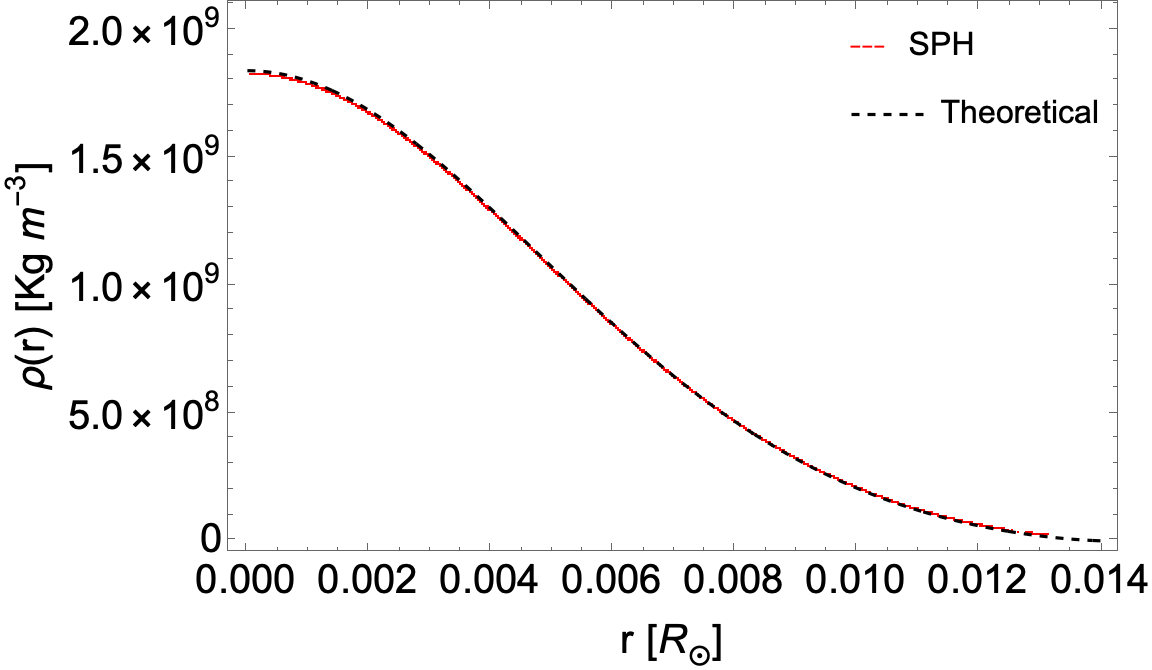}
	\caption{Density, $\rho(r)$ vs radial distance, $r$. \textbf{Left Panel: }WD configuration: $M_{\mathrm{WD}}=0.25 M_{\odot}$ and $R_{\mathrm{WD}}=0.0193 R_{\odot}$. \textbf{Right Panel: }WD configuration: $M_{\mathrm{WD}}=0.5 M_{\odot}$ and $R_{\mathrm{WD}}= 0.0141 R_{\odot}$}
	\label{rho_profile1}
\end{figure}

\begin{figure}[h]
	\centering
	\includegraphics[width=.45\textwidth]{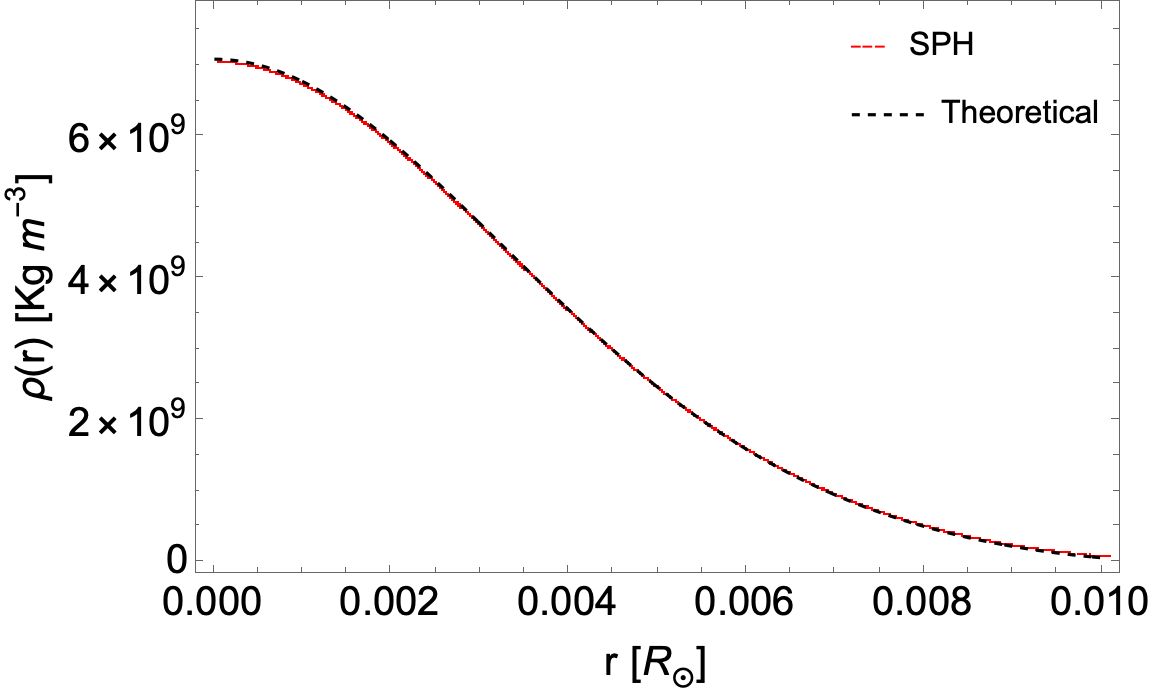}
	\hfill
	\includegraphics[width=.45\textwidth]{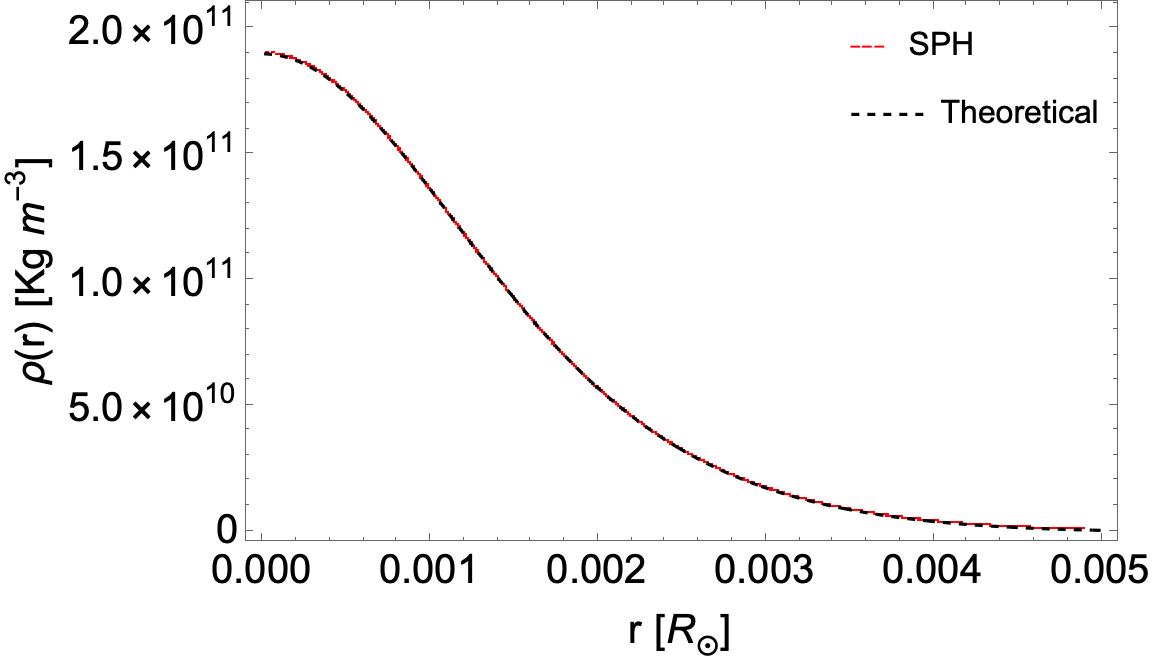}
	\caption{Density, $\rho(r)$ vs radial distance, $r$. \textbf{Left Panel: }WD configuration: $M_{\mathrm{WD}}=0.75 M_{\odot}$ and $R_{\mathrm{WD}}=0.0109 R_{\odot}$. \textbf{Right Panel: }WD configuration: $M_{\mathrm{WD}}=1.25 M_{\odot}$ and $R_{\mathrm{WD}}= 0.0053 R_{\odot}$}
	\label{rho_profile2}
\end{figure}

\purple{Correspondingly, in Figures \ref{rho_profile1} and \ref{rho_profile2}, we confirm a few SPH internal profile with their respective theorectical solution of $\rho(r)$ (or $x(r)$). Once, these profiles of $x(r)$ for the relaxed SPH WDs are matched, consequently most other properties are also verified. Since, quantities such as pressure, internal energy, and related variables fundamentally depend on $x(r)$. Therefore, the relaxed configuration of a considered WD can found to be in agreement with Virial equilibrium.}

\section{Convergence of SPH Results}\label{app:B}
\begingroup
\color{black}
To ensure the consistency of our SPH results, we verified the key findings by performing simulations with a higher SPH resolution, using $N_{\mathrm{SPH}} = 10^6$ particles. 

It is sufficient to focus on the more general results obtained in our analysis, namely the TDEs of spinning WDs. In particular, the off-equatorial configuration ($\theta_a = 1^\circ$, $a^{\star} = \pm 0.98$) is crucial, as it reveals degeneracy splitting in the outcomes of the variables and observables. These resolution checks are presented in the left panels of Figures \ref{res:core_mass}, \ref{res:kick_vel}, and \ref{res:fallback}, for the core mass fraction, kick velocity, and fallback rates, respectively. The right panels of these figures show the corresponding resolution checks for the case of a non-spinning WD with $a^{\star} = -0.98$ and $\theta_a = 1^\circ$.

\begin{figure}[h]
	\centering
	\includegraphics[width=.45\textwidth]{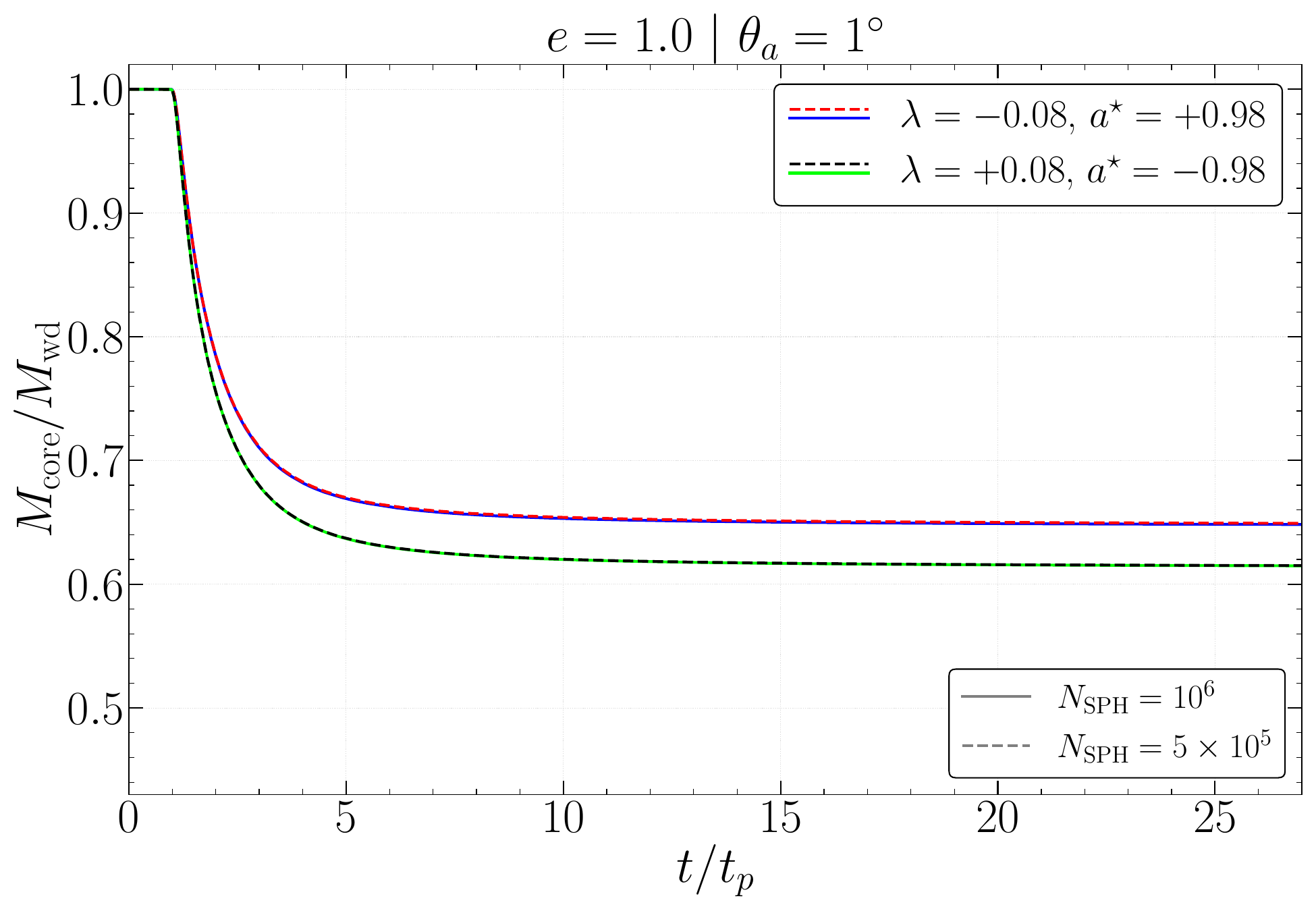}
	\hfill
	\includegraphics[width=.45\textwidth]{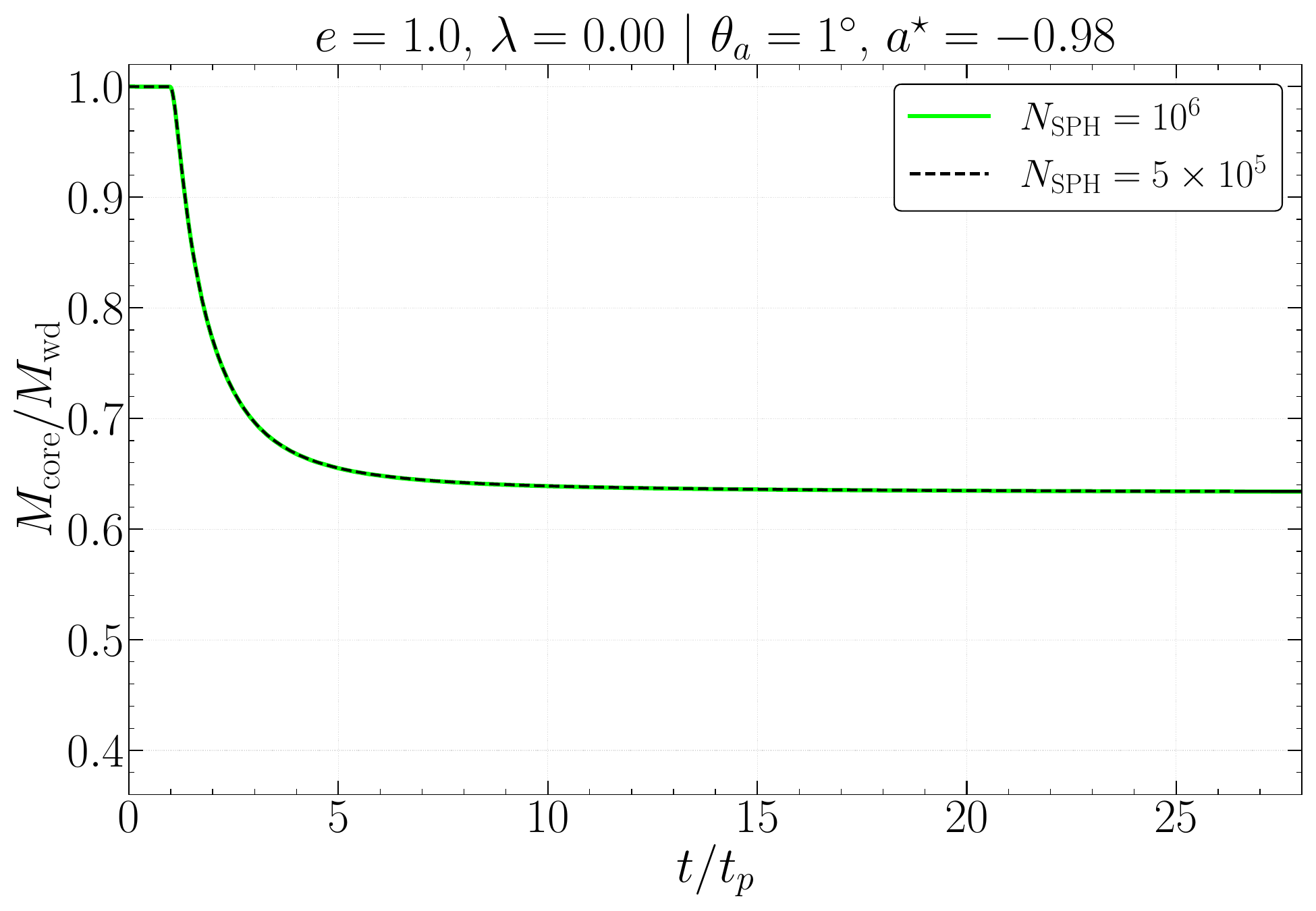}
	\caption{Resolution check for core mass fraction with $N_{\mathrm{SPH}}=10^6$ (denoted by solid lines) and $N_{\mathrm{SPH}}=5\times 10^5$ (denoted by dashed lines). \textbf{Left Panel:}  The results corresponds to $\lambda=-0.08$ (with $a^\star=+0.98$), denoted by dashed red-solid blue lines and $\lambda=+0.08$ (with $a^\star=-0.98$), denoted by dashed black-solid green lines.  \textbf{Right Panel: }Results corresponding to non-spinning WD, $\lambda = 0.0$. The result is shown for $a^\star=-0.98$ case, denoted by dashed black-solid green lines.}
	\label{res:core_mass}
\end{figure}

\begin{figure}[h]
	\centering
	\includegraphics[width=.45\textwidth]{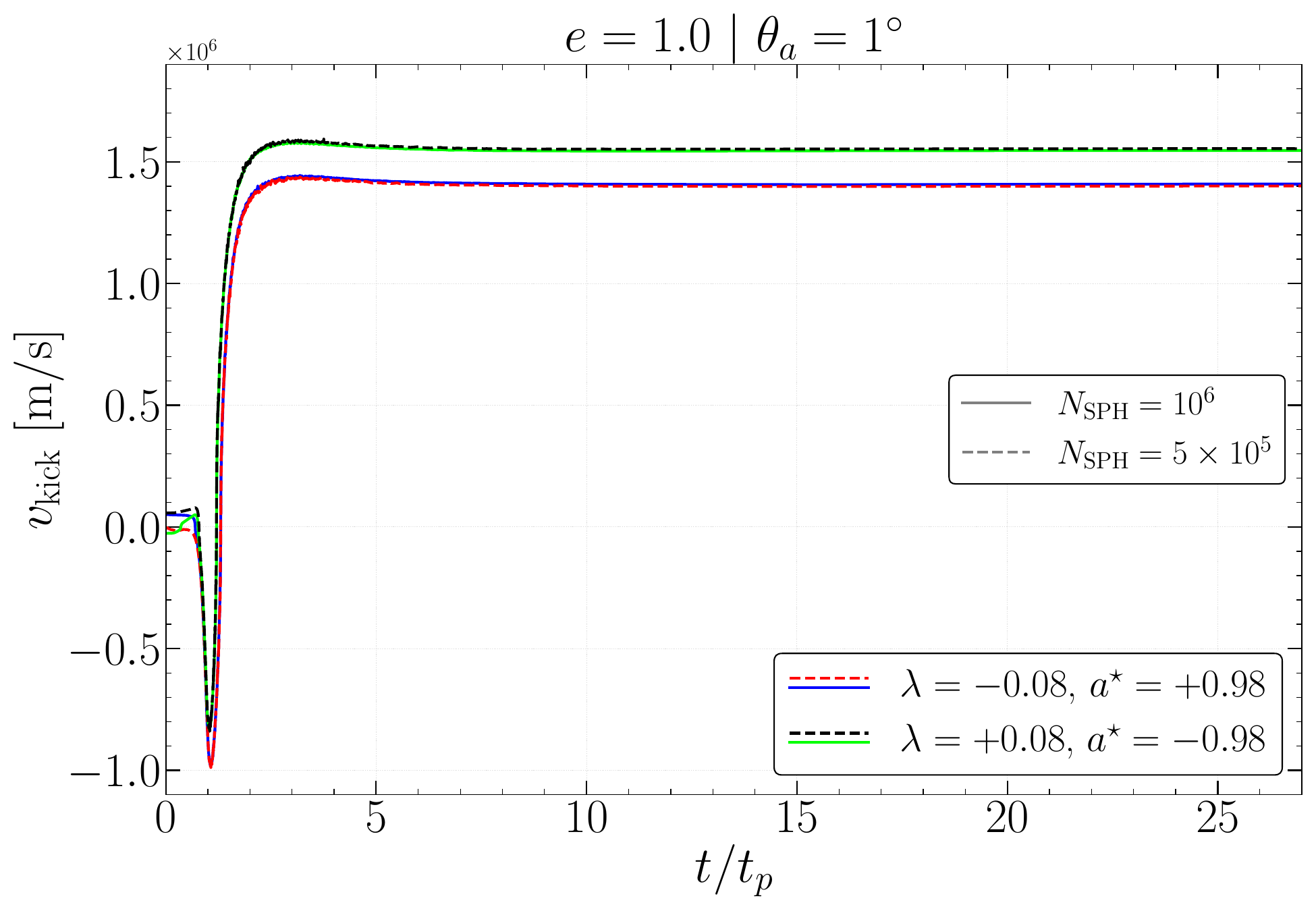}
	\hfill
	\includegraphics[width=.45\textwidth]{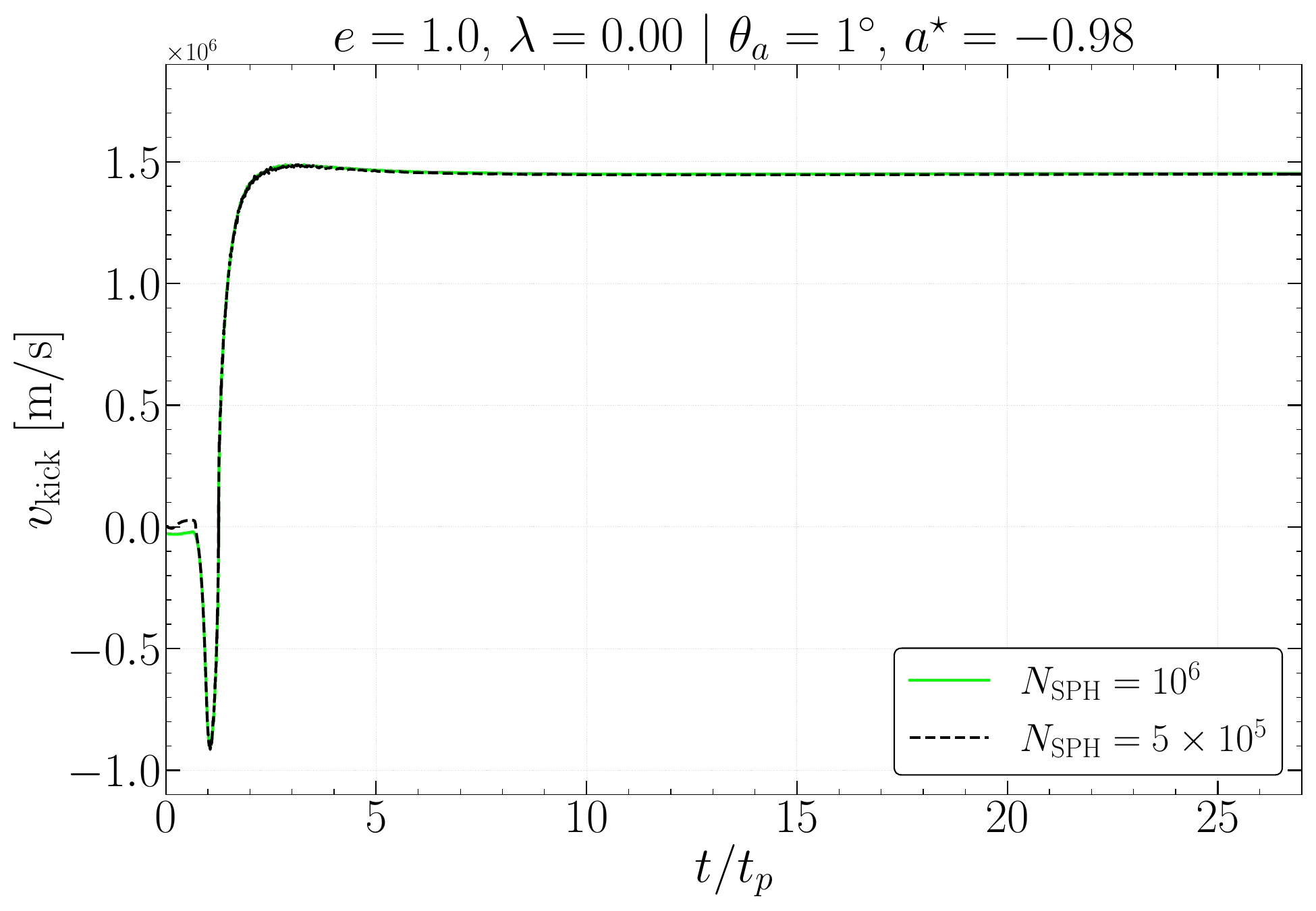}
	\caption{Resolution check for kick velocity with $N_{\mathrm{SPH}}=10^6$ (denoted by solid lines) and $N_{\mathrm{SPH}}=5\times 10^5$ (denoted by dashed lines). \textbf{Left Panel:} Results corresponding to spinning WD, $\lambda\ne 0.0$. \textbf{Right Panel: }Results corresponding to non-spinning WD, $\lambda = 0.0$.}
	\label{res:kick_vel}
\end{figure}

\begin{figure}[h]
	\centering
	\includegraphics[width=.45\textwidth]{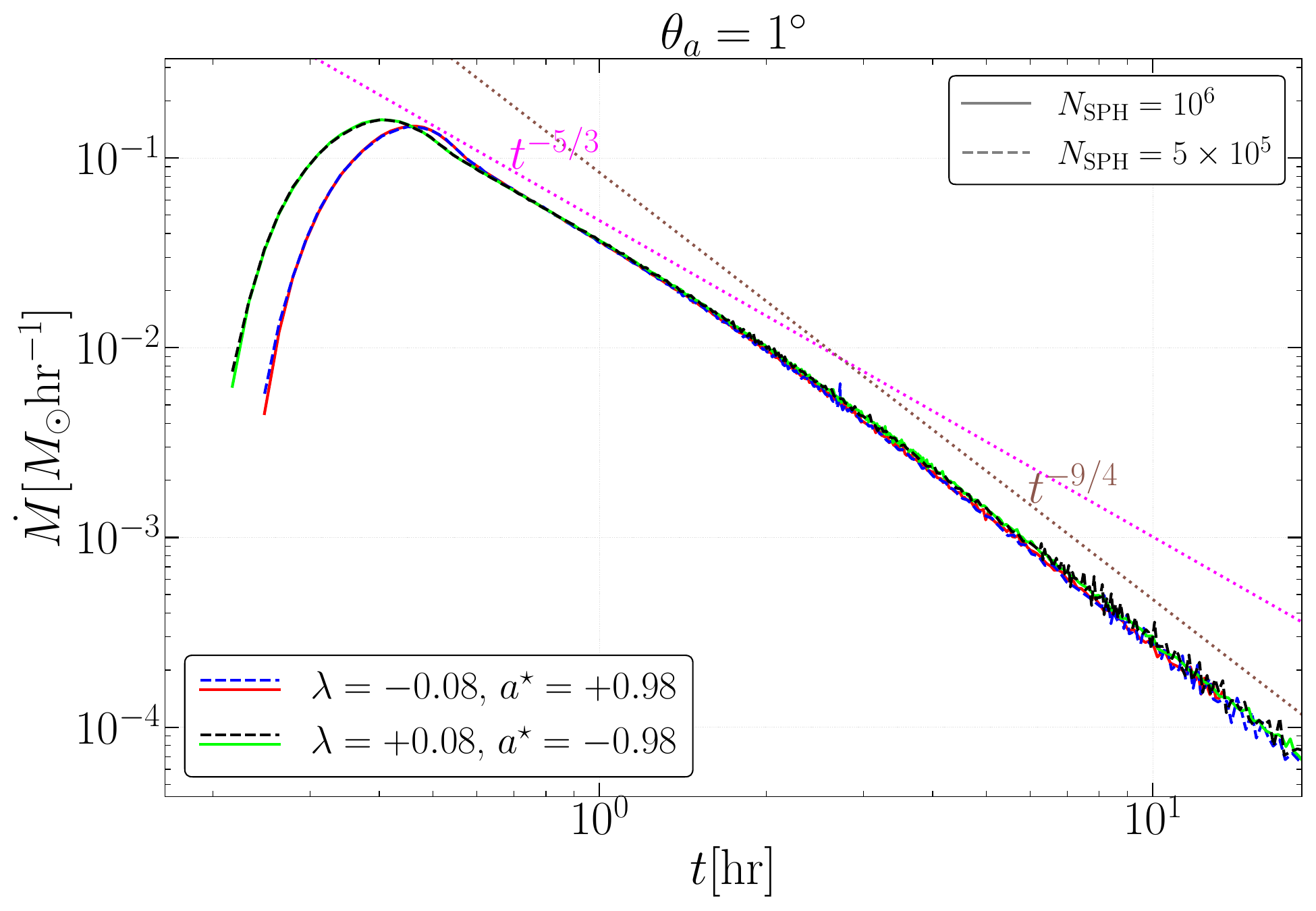}
	\hfill
	\includegraphics[width=.45\textwidth]{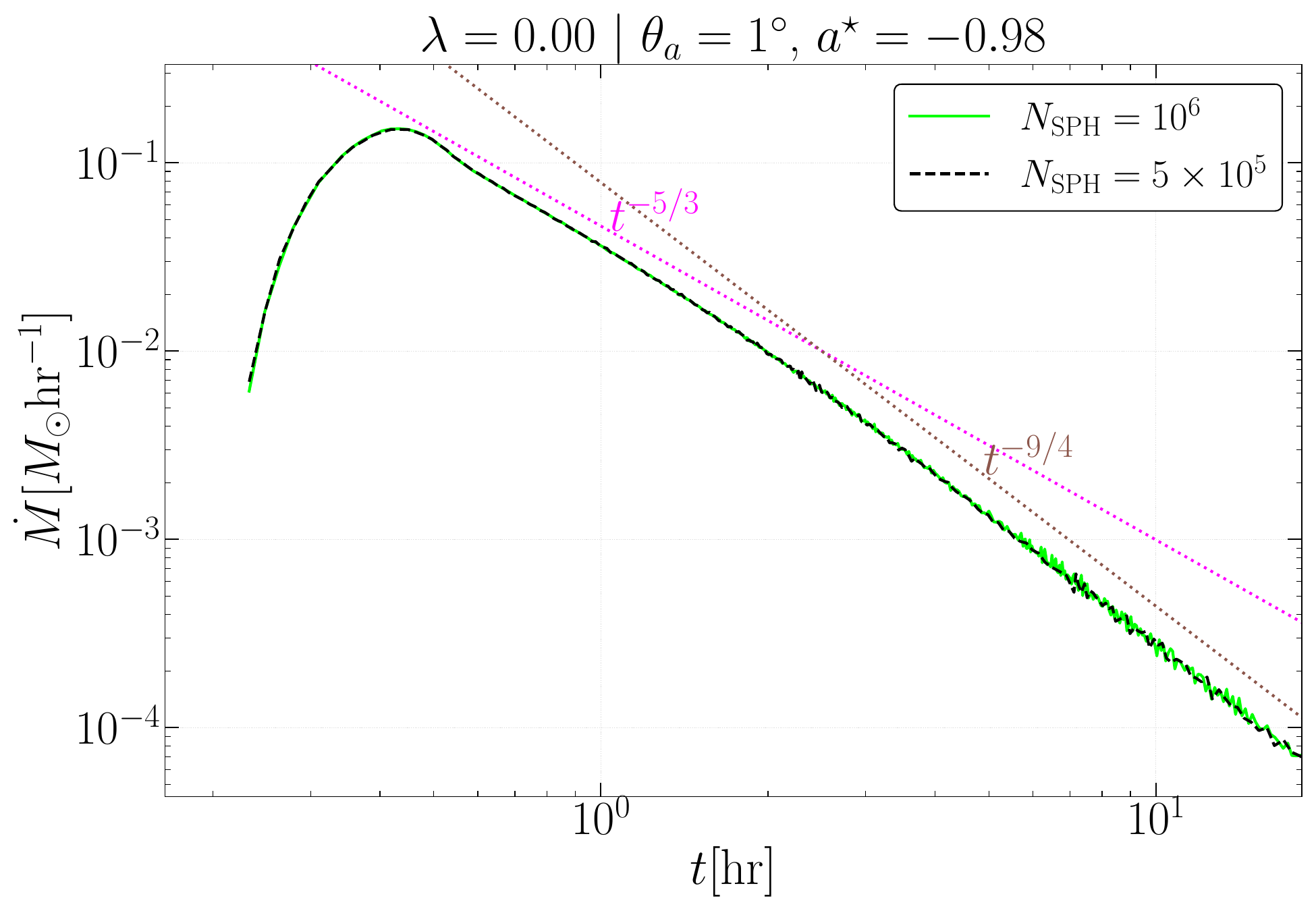}
	\caption{Resolution check for fallback rates with $N_{\mathrm{SPH}}=10^6$ (denoted by solid lines) and $N_{\mathrm{SPH}}=5\times 10^5$ (denoted by dashed lines). \textbf{Left Panel:} Results corresponding to spinning WD, $\lambda\ne 0.0$. \textbf{Right Panel: }Results corresponding to non-spinning WD, $\lambda = 0.0$.}
	\label{res:fallback}
\end{figure}

The SPH results obtained with $N_{\mathrm{SPH}} = 5 \times 10^5$ and $N_{\mathrm{SPH}} = 10^6$ are in good agreement, demonstrating full convergence. Moreover, before presenting our final results, we also performed simulations with $N_{\mathrm{SPH}}=10^5$, where the outcomes had already converged. Therefore, the results presented in this work are independent of higher SPH resolution.
\endgroup

\section*{Data Availability Statement}
The data underlying this article will be shared upon reasonable request to the corresponding author.

\providecommand{\href}[2]{#2}\begingroup\raggedright
\endgroup

\end{document}